\documentclass[reprint, aps, pra, twoside, floatfix, nofootinbib, longbibliography]{revtex4-2}
\usepackage{amsmath, amssymb}
\usepackage{graphicx}
\usepackage[usenames, dvipsnames]{color}
\usepackage{mathtools}
\usepackage{hyperref}
\hypersetup{colorlinks=true, linkcolor = blue, citecolor = blue, urlcolor = blue}
\usepackage{xcolor}
\usepackage{bbold}
\usepackage{orcidlink}

\newcommand{\opU}{{\mathcal{U}}}
\newcommand{\opH}{{\mathcal{H}}}
\newcommand{\opbH}{{\bar{\mathcal{H}}}}

\newcommand{\opO}{\mathcal{O}}
\newcommand{\opZ}{\mathcal{Z}}
\newcommand{\mathhc}{\textrm{H.c.}}

\newcommand{\dd}{\mathrm{d}}
\newcommand{\ee}{\mathrm{e}}
\newcommand{\ii}{\mathrm{i}}
\newcommand{\pdt}{\frac{\partial}{\partial t}}
\newcommand{\vacket}{|\varnothing\rangle}
\newcommand{\vacbra}{\langle\varnothing|}
\newcommand{\gnd}{\mathcal{G}}
\newcommand{\pol}{\mathcal{E}}
\newcommand{\omegaa}{\omega_a}
\newcommand{\omegab}{\omega_b}
\newcommand{\NA}{{\mathcal{N}}_a}
\newcommand{\NB}{{\mathcal{N}}_b}
\newcommand{\res}{\mathrm{r}}
\newcommand{\wg}{\mathrm{w}}
\newcommand{\tmn}{\mathrm{t}}
\newcommand{\sys}{\mathrm{sys}}
\newcommand{\Heff}{\mathcal{H}_\mathrm{eff}}
\everymath{\displaystyle}
\begin{document}
\title{Generation of time-frequency entangled photon pairs propagating in separate waveguides in circuit QED setup}
\author{E.~V.~Stolyarov\,\orcidlink{0000-0001-7531-5953}}
\email{eugenestolyarov@gmail.com}
\affiliation{Bogolyubov Institute for Theoretical Physics, National Academy of Sciences of Ukraine, Metrolohichna Street 14-b, 03143 Kyiv, Ukraine}
\begin{abstract}
Time-frequency entangled photons constitute an important resource for a plethora of applications across the diverse quantum technology landscape.
Thus, efficient and tunable setups for the generation of entangled photons are requisite for modern quantum technologies.
In this work, we propose a generic cavity QED setup designed for on-demand generation of time-frequency entangled photon pairs, with each photon propagating in a separate waveguide. We outline a potential incarnation of this setup in the microwave superconducting circuit QED architecture.
We derive and numerically solve the set of equations of motion governing the evolution of the quantum state of the system, allowing us to examine the photon emission dynamics.
Using the Schmidt decomposition of the joint spectral amplitude of the emitted photon pair, we compute the entanglement entropy analyzing its dependence on the system parameters.
We outline the potential extension of the proposed scheme for the generation of multiphoton time-frequency entangled states.
\end{abstract}
\setcounter{page}{1}
\maketitle
\section{Introduction} \label{sec:intro}
The photonic time-frequency (TF) degree of freedom can serve as a basis for encoding quantum information, including time \cite{kurp2019}, frequency \cite{lukens2017}, and temporal-mode \cite{brecht2015,*ansari2018} encodings.
Moreover, the TF degree is intrinsically robust for long-distance transmission of quantum information via microwave and optical waveguides.
Nonclassical spectral and temporal correlations of photonic TF entangled states can be harnessed for boosting the sensitivity and resolution of the two-photon absorption spectroscopy \cite{schlawin2018, geaban1989, *jav1990, *tabakaev2021, oka2018, *oka2020} and ultrafast TF-resolved Raman spectroscopy \cite{zhang2022} and enhancing the detection efficiency of low-reflectivity objects in the bright thermal environment \cite{lloyd2008, barz2015, *barz2020}.
The large dimensionality of TF entangled states is instrumental in the realization of high-dimensional quantum key distribution protocols \cite{tittel2000, alikhan2007, *nunn2013}.
Besides, TF entangled states proved to be useful in a range of other applications, such as testing Bell inequalities \cite{franson1989, tittel1998, cabello2009}, quantum lithography \cite{dangelo2001}, quantum-enhanced clock synchronization \cite{giovannetti2001}, and biomedical imaging \cite{varnavski2022}.
However, most of the experimental implementations of these quantum protocols and techniques use infrared and visible-light photons.
Having an efficient and scalable scheme for the generation of TF entangled photons in the microwave domain, one can leverage that diverse quantum-optical toolbox in the elaborate microwave photonic platform based on superconducting quantum circuits \cite{gu2017}.
Rapid progress and ongoing advances in the development of microwave superconducting circuit QED systems demonstrated the high potential of these quantum systems as a versatile hardware architecture for the realization of quantum information processing (QIP) devices \cite{wendin2017, *blais2020}.

A number of theoretical proposals \cite{marquardt2007, ychang2016, sburillo2016, sathya2016} and experimental demonstrations \cite{lepp2012, westig2017, peugeot2021, flurin2012, gaspar2017, wren2020, perel2022} of various designs of deterministic and probabilistic circuit QED sources of TF entangled microwave photons were put forward.
These systems exploit the nonlinear properties of the Josephson junctions.
One of the approaches consists in using voltage-biased Josephson junctions interacting with modes of open microwave resonators.
Due to strong nonlinear light-charge interaction, the inelastic tunneling of Cooper pairs across the Josephson junction can generate nonclassical states of light \cite{lepp2012, westig2017, peugeot2021, kubala2015, rolland2019, *grimm2019, ma2021, *ma2022}, particularly TF entangled photons \cite{lepp2012, westig2017, peugeot2021}.
Alternatively, one can employ the nondegenerate Josephson mixer -- a circuit QED device, which parametrically couples two modes with different frequencies by a pump at their sum-frequency \cite{roch2012, bergeal2012}.
In Ref.~\cite{flurin2012}, the authors demonstrated the generation of entangled microwave radiation over two separate transmission lines using the nondegenerate Josephson mixer.
A Josephson metamaterial operating in a Kerr-free three-wave mixing mode was utilized for the generation of TF entangled microwave photons in Ref.~\cite{perel2022}.
Another approach is leveraging superconducting artificial atoms, which are anharmonic multilevel quantum systems \cite{clarke2008, kjaer2020}.
By populating higher excited states of an artificial atom via driving the two-photon transition, one can trigger the process of spontaneous cascaded relaxation leading to the emission of TF entangled photons \cite{marquardt2007, ychang2016, sathya2016, sburillo2016, gaspar2017, wren2020}.

\begin{figure*}[t!] 
	\centering
	\includegraphics{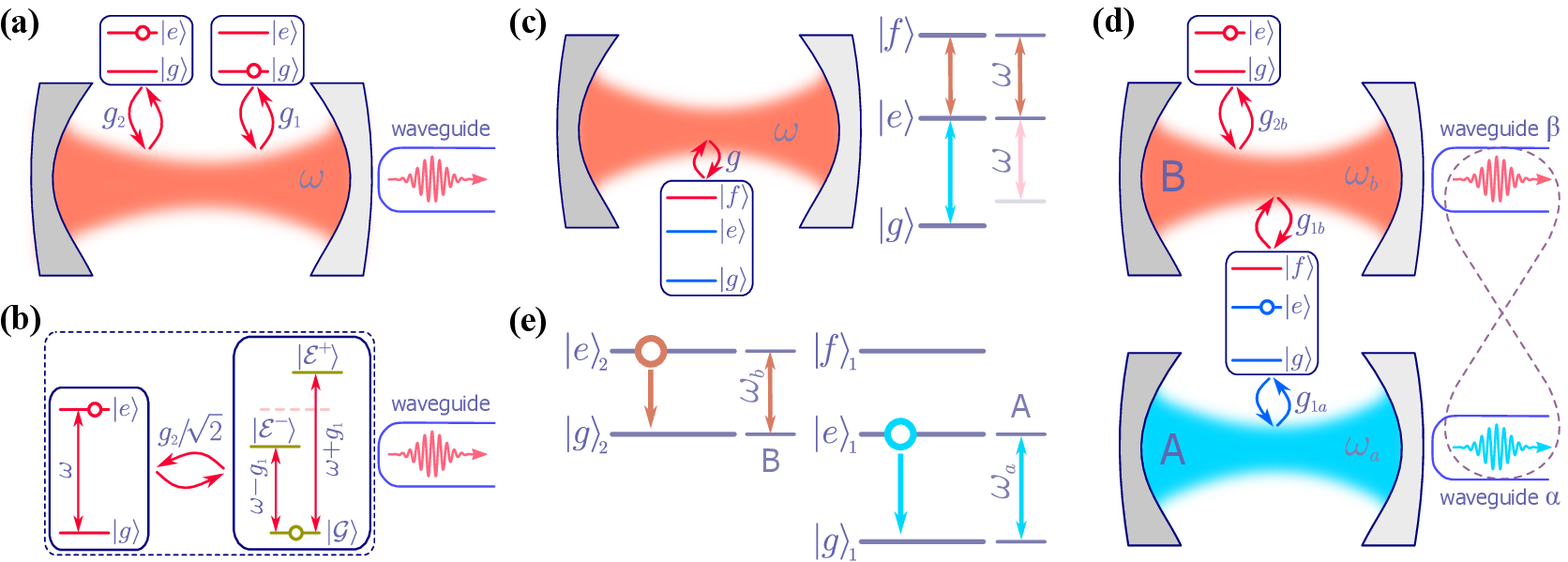}
	\caption{(a) Schematic representation of the system comprised of a pair of 2LEs coupled to a mode of an open resonator.
		(b) The equivalent representation of the system depicted in Fig.~\hyperref[fig:fig_1]{\ref*{fig:fig_1}(a)} -- a 2LE interacting with a $\textsf{V}$3LE coupled to a waveguide.
		(c) Scheme of the system composed of a resonator interacting with a $\textstyle \Xi$-configuration 3LE and the diagram showing the interrelation between the $\textstyle \Xi$3LE transition frequencies and the resonator frequency.
		(d) The generic schematics of the proposed source of TF entangled photon pairs.
		(e) Diagram of the interrelation between the frequencies of the resonators and the transition frequencies of the emitters in the setup demonstrated in Fig.~\hyperref[fig:fig_1]{\ref*{fig:fig_1}(d)}. \label{fig:fig_1}}
\end{figure*}

In the paper, we start with proposing a general scheme of the setup designed for the generation of TF entangled photons and then outline its potential experimental implementation as a microwave photonic device based on a superconducting circuit QED platform.
In this scheme, the correlated relaxation of a pair of excited quantum emitters is realized when the relaxation of one of the emitters into the open resonator triggers the relaxation of its counterpart leading to the emission of a pair of TF entangled photons.
One of the advantages of the proposed scheme for generation of TF entangled photon pairs is that it does not require driving the two-photon transitions of the artificial atoms.
Moreover, it can be extended for the generation of multiphoton entangled states.
We focus on studying the proposed device in the pulsed on-demand regime, when exactly only one pair of entangled photons are emitted. 
Using a state-vector approach, we provide a real-time picture of photon emission dynamics.
We use the Schmidt decomposition of the joint spectral amplitude of the emitted photons for computing the von Neumann (entanglement) entropy, which is employed as a measure of their bipartite entanglement.

The structure of the paper is as follows.
In Sec.~\ref{sec:scheme}, we discuss the general principles determining the operation of the proposed source of TF entangled photon pairs.
Its potential implementation within the microwave circuit QED architecture is outlined in Sec.~\ref{sec:cirqed}.
In Sec.~\ref{sec:model}, we present the model Hamiltonian describing the system.
In Sec.~\ref{sec:evol}, we derive the equations of motion governing the quantum-state evolution of the system and study the emission dynamics.
The dependence of the entanglement degree of the emitted photons on the system parameters is analyzed in Sec.~\ref{sec:entang}.
We summarize our results and outline possible applications and extensions of the proposed source of entangled photon pairs in Sec.~\ref{sec:concl}.
The additional considerations and details of derivations are delegated to the Appendixes.

\section{General scheme and operational principle} \label{sec:scheme}
To explain the key principles defining the operation of the proposed source of TF entangled photon pairs, let us consider a paradigmatic system schematically presented in Fig.~\hyperref[fig:fig_1]{\ref*{fig:fig_1}(a)}.
This system consists of an open resonator interacting with a pair of two-level emitters (2LEs).
The first 2LE is coupled to the resonator with strength $\textstyle g_1$, and the second 2LE interacts with the resonator with strength $\textstyle g_2$.
There is no direct interaction between the emitters.
For simplicity, we assume that the resonator and the 2LEs have identical frequencies $\textstyle \omega$.
The resonator is open due to its coupling to a waveguide, which leads to the photon leakage from the resonator with rate $\textstyle \kappa$.
Initially, the first 2LE resides in the ground state, while the second 2LE is prepared in the excited state.
Due to interaction with the open resonator, the second 2LE eventually decays, delivering the excitation (photon) to the waveguide.

The system, shown in Fig.~\hyperref[fig:fig_1]{\ref*{fig:fig_1}(a)}, can be represented as the excited-state 2LE interacting with the $\textstyle \textsf{V}$-configuration three-level emitter (3LE) coupled to the waveguide.
Figure~\hyperref[fig:fig_1]{\ref*{fig:fig_1}(b)} illustrates the equivalent representation, whose detailed justification is provided in Appendix~\ref{sec:appa}.
The $\textstyle \textsf{V}$-type 3LE is formed by the ground state $\textstyle |\gnd\rangle$ and a pair of excited states $\textstyle |\pol^\pm\rangle$.
The latter decay to the waveguide with the rate $\textstyle \kappa/2$.
The frequency of $\textstyle |\gnd\rangle\leftrightarrow|\pol^\pm\rangle$ transition is $\textstyle \omega \pm g_1$.
Both transitions are coupled to the 2LE with strength $\textstyle g_2/\sqrt{2}$.
The increase of coupling $\textstyle g_1$ leads to stronger detuning of $\textstyle |\gnd\rangle\leftrightarrow|\pol^\pm\rangle$ transition from the 2LE frequency $\textstyle \omega$.
For $\textstyle g_1 > g_2/\sqrt{2}$, the excitation exchange between the 2LE and the $\textstyle \textsf{V}$3LE becomes inefficient, resulting in a suppression of the photon leakage into the waveguide.
These qualitative considerations are confirmed by the calculations revealing the suppression of the photon leakage into the waveguide with the increase of the ratio $\textstyle g_1/g_2$.
Details of calculations are given in Appendix~\ref{sec:appa}.

The above result implies that the process of photon emission can be manipulated by controlling the coupling parameter $\textstyle g_1$.
In this regard, let us consider a system comprised by a 3LE interacting with a single-mode resonator.
This system is sketched in Fig.~\hyperref[fig:fig_1]{\ref*{fig:fig_1}(c)}.
The ground state $\textstyle |g\rangle$ and the excited states $\textstyle |e\rangle$ and $\textstyle |f\rangle$ of the 3LE are arranged in a ladder ($\textstyle \Xi$-type) configuration implying that only $\textstyle |g\rangle \leftrightarrow |e\rangle$ and $\textstyle |e\rangle \leftrightarrow |f\rangle$ transitions are allowed.
The frequencies of the resonator and $\textstyle |e\rangle \leftrightarrow |f\rangle$ transition coincide, while $\textstyle |g\rangle \leftrightarrow |e\rangle$ transition is strongly detuned from the resonator.
In this case, the interaction between the 3LE and the resonator is approximately described by the Hamiltonian $\textstyle \opH_\mathrm{int} = g \, (c^\dag |e\rangle\langle f| + |f\rangle\langle e|c)$, where $\textstyle g$ denotes the coupling parameter, and the operator $\textstyle c$ ($\textstyle c^\dag$) annihilates (creates) a resonator photon.
One can formally rewrite the resonator-emitter interaction Hamiltonian as $\textstyle \opH_\mathrm{int} = g(|e\rangle\langle e| + |f\rangle\langle f|) (c^\dag |e\rangle\langle f| + |f\rangle\langle e|c)$.
Using the property $\textstyle |g\rangle\langle g|+|e\rangle\langle e|+|f\rangle\langle f| = \mathbb{1}$, with $\textstyle \mathbb{1}$ being the unity operator, leads to
$\textstyle \opH_\mathrm{int} = \tilde{g} (c^\dag |e\rangle\langle f| + |f\rangle\langle e|c)$, where $\textstyle \tilde{g} \equiv g(\mathbb{1}-|g\rangle\langle g|)$ can be interpreted as a resonator-emitter coupling dependent on the $\textstyle \Xi$3LE ground state population.
Thus, if the 3LE resides in the ground state $\textstyle |g\rangle$, its $\textstyle |f\rangle\leftrightarrow|e\rangle$ transition is decoupled from the resonator mode.
This effect was utilized, e.g., in Ref.~\cite{kyr2016} for the implementation of a continuous-wave single-photon transistor and Ref.~\cite{sto2020} for the implementation of a qubit-state-controlled single-photon switch.

Harnessing the effects outlined above, we propose a setup consisting of two emitters interacting with two resonators (marked as $\textstyle A$ and $\textstyle B$) coupled to the output waveguides (marked as $\textstyle \alpha$ and $\textstyle \beta$).
The considered setup is schematically shown in Fig.~\hyperref[fig:fig_1]{\ref*{fig:fig_1}(e)}.
In this setup, the first emitter is coupled to both resonators.
This emitter is represented by a $\textstyle \Xi$-configuration 3LE with the ground state $\textstyle |g\rangle_1$ and a pair of excited states $\textstyle |e\rangle_1$ and $\textstyle |f\rangle_1$.
The frequency of $\textstyle |g\rangle_1 \leftrightarrow |e\rangle_1$ transition coincides with the frequency $\textstyle \omega_a$ of resonator $\textstyle A$, while the frequency of $\textstyle |e\rangle_1 \leftrightarrow |f\rangle_1$ transition coincides with the frequency $\textstyle \omega_b$ of resonator $\textstyle B$.
The anharmonicity of the eigenlevels of the $\textstyle \Xi$3LE results in the inhibition of the excitation exchange between resonator $A$ and $\textstyle |e\rangle_1 \leftrightarrow |f\rangle_1$ transition and between resonator $\textstyle B$ and $\textstyle |g\rangle_1 \leftrightarrow |e\rangle_1$ transition.
The second emitter is represented by a 2LE with the ground state $\textstyle |g\rangle_2$ and the excited state $\textstyle |e\rangle_2$.
This emitter is coupled only to resonator $\textstyle B$, and 
the frequency of its $\textstyle |g\rangle_2 \leftrightarrow |e\rangle_2$ transition coincides with the resonator frequency $\textstyle \omegab$.
The interrelation between the frequencies of the resonators and the transitions frequencies of the emitters is shown in Fig.~\hyperref[fig:fig_1]{\ref*{fig:fig_1}(d)}.

In the system outlined above, the \emph{correlated relaxation} of the emitters emerges.
The mechanism of this relaxation is as follows.
When the 3LE resides in its excited state $\textstyle |e\rangle_1$, the relaxation of the 2LE into waveguide $\textstyle \beta$ via resonator $\textstyle B$ is inhibited due to the mechanism outlined in the second paragraph of this section.
In the course of the 3LE relaxation from the excited state $\textstyle |e\rangle_1$ to the ground state $\textstyle |g\rangle_1$, $\textstyle |f\rangle_1 \leftrightarrow |e\rangle_1$ transition of the 3LE is getting decoupled from the resonator $\textstyle B$, as discussed earlier in this section, which triggers the relaxation of the 2LE excited state $\textstyle |e\rangle_2$.
Hence, the photons tend to be emitted into the waveguides one right after another, implying that they exhibit \emph{time correlation} which, in turn, indicates their \emph{frequency (energy) anti-correlation}.

\begin{figure}[t!]
	\centering
	\includegraphics{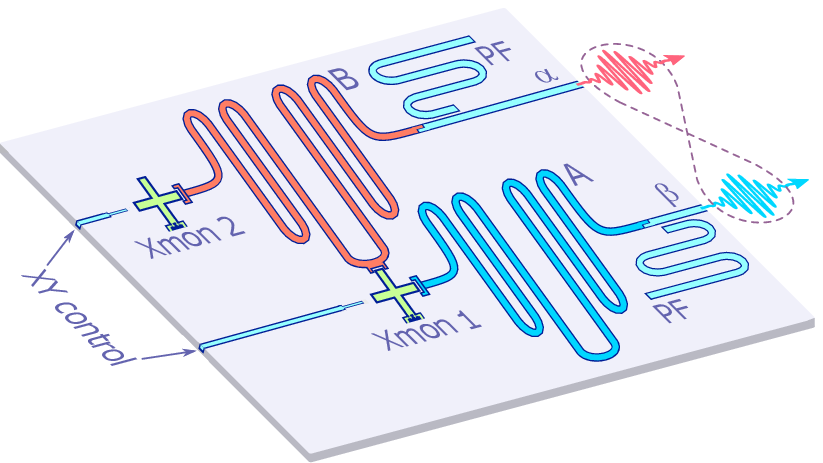}
	\caption{Sketch of the potential experimental circuit QED implementation of the generic scheme illustrated in Fig.~\hyperref[fig:fig_1]{\ref*{fig:fig_1}(e)}.
		The indices of the elements are in correspondence to the scheme in Fig.~\hyperref[fig:fig_1]{\ref*{fig:fig_1}(e)}.
		Each output transmission line is additionally equipped with the Purcell filter (PF).
		The excited states of the artificial atoms (Xmons) are individually prepared using the classical pulses sent over XY control lines. \label{fig:fig_2}}
\end{figure}

\section{Circuit QED setup} \label{sec:cirqed}
Having discussed the general principle of operation of the proposed source of TF entangled photon pairs, let us now outline its potential implementation in the circuit QED architecture.
The superconducting circuit QED platform provides all necessary components for implementation of the generic scheme discussed in Sec.~\ref{sec:scheme}, such as microwave resonators and transmission lines, as well as Josephson-junction artificial atoms featuring tunable multilevel structure and long coherence times \cite{kjaer2020}.

The schematic illustration of the proposed circuit QED setup is shown in Fig.~\ref{fig:fig_2}.
In the considered setup, the quantum emitters are represented by the Xmon version \cite{barends2013} of the transmon -- a charge-type artificial atom derived from a Cooper-pair box and featuring resilience to charge fluctuations \cite{koch2007, *houck2009}.
Transmon artificial atoms, in particular, their Xmon modifications offer a simple design, long coherence times, and tunable parameters \cite{yuchen2014}, making them a popular choice for the realization of various QIP systems \cite{barends2014, kelly2015, riste2017}.
Moreover, using tantalum as a base superconductor along with improving processing techniques led to a recent impressive increase in the transmon excited state lifetime \cite{place2021, wang2022}.
In our setup, the Xmons are capacitively coupled to the coplanar waveguide (CPW) resonators which, in turn, are coupled to the transmission-line waveguides serving as output channels for the emitted photons.

The states of the transmons are manipulated using the classical signals delivered via individual (XY) control lines \cite{krantz2019}.
Due to the weak anharmonicity of transmons, for their preparation in the lower excited state, one can rely on one of the elaborate techniques, providing high fidelities of preparation while avoiding excitation of the higher excited states \cite{motzoi2009, chow2010, machnes2018}.
Alternatively, for transmon control, one can use a single-flux quantum digital logic \cite{leonard2019}.
This approach allows integration of the control circuitry on a chip bringing most of the setup components into the cryogenic stage.

The output transmission lines are supplemented with the Purcell filters for suppression of the superfluous emission from the off-resonant transitions of the transmons (more details on the role of these elements are given in Appendix~\ref{sec:purcell}).
The setup presented in Fig.~\ref{fig:fig_2} adopts the approach from Ref.~\cite{reed2010} with the quarter-wavelength CPW resonators side-coupled to the output transmission lines.
These additional resonators serve as notch filters.

In the setup version shown in Fig.~\ref{fig:fig_2}, the fixed couplings between the circuit elements are used.
For achieving \emph{in-situ} tunability of the setup parameters for better control over the characteristics of the emitted photons, instead of fixed couplings set on a chip fabrication stage, one can implement tunable couplings using, for example, superconducting quantum interference device couplers \cite{peropadre2013, koun2018, *wu2018} controlled by an external flux.

\section{The Model} \label{sec:model}
The circuit QED system described above is modeled by the Hamiltonian
\begin{equation} \label{eq:ham_1}
 \begin{split}
   \opH_\sys = & \opH_\tmn + \opH_\mathrm{r} + \opH_{\res-\tmn} + \opH_\wg + \opH_{\res-\wg},
 \end{split}
\end{equation}
where the first term describes the transmon-type artificial atoms, the second term is the Hamiltonian of the resonators, the third term describes the interaction between the resonators and transmons, and the last pair of terms describes the waveguides and their coupling to the resonators.
In what follows we set $\textstyle \hbar = 1$, thus, measuring all energies in frequency units.

The Hamiltonian $\textstyle \opH_\tmn$ represented in the basis of the transmons eigenstates reads as
\begin{equation} \label{eq:ham_tmn2}
	\opH_\tmn = \sum_{j=1}^2 \left(\omega_j^{ge} \sigma^{ee}_j	+ \omega^{gf}_j \sigma^{ff}_j\right),
\end{equation}
where $\textstyle \omega^{ge}_j$, $\textstyle \omega^{gf}_j = \omega_j^{ge} + \omega_j^{ef}$, and $\omega^{ef}_j$ stand for the frequency of $\textstyle |g\rangle_j \leftrightarrow |e\rangle_j$, $\textstyle |g\rangle_j \leftrightarrow |f\rangle_j$, and $\textstyle |e\rangle_j \leftrightarrow |f\rangle_j$ transition of the $\textstyle j$-th transmon ($\textstyle j\in\{1,2\}$), respectively.
For convenience, in Eq.~\eqref{eq:ham_tmn2} we introduced the transmon operator $\textstyle \sigma_j^{ss'} = |s\rangle_j \langle s'|_j$ where $\textstyle s,s' \in \{g, e, f\}$ is the index of the transmon eigenlevel.
Since we consider the case of only two excitations in the system, we restricted the subspace of the transmon states to the ground state $\textstyle |g\rangle_j$ and a pair of the lowest excited states $\textstyle |e\rangle_j$ and $\textstyle |f\rangle_j$.

The Hamiltonian of the resonators $\textstyle \opH_\res$ has the form
\begin{equation}
  \opH_\res = \omegaa \NA + \omegab \NB,
\end{equation}
where $\textstyle \omegaa$ and $\textstyle \omegab$ denote the frequencies of resonator $\textstyle A$ and $\textstyle B$, respectively.  
Operators $\textstyle \NA = a^\dag a$ and $\textstyle \NB = b^\dag b$ are the operators of the photon number in resonator $\textstyle A$ and $\textstyle B$, where $\textstyle a$($\textstyle a^\dag$) and $\textstyle b$ ($\textstyle b^\dag$) stand for the photon annihilation (creation) operators in the corresponding resonator.

For clarity, we represent the resonator-transmons interaction Hamiltonian as $\textstyle \opH_{\res-\tmn} = \opH^\mathrm{rsn}_{\res-\tmn} + \opH^\mathrm{dsp}_{\res-\tmn}$, where the first term describes the \emph{resonant} interactions determining the dynamics of the system, while the second term describes the \emph{dispersive} interactions giving rise to shifts in the frequencies of the resonators and the transmons.
In the rotating-wave approximation, the term $\textstyle \opH^\mathrm{res}_\mathrm{r-a}$ has the form
\begin{equation} \label{eq:ham_rq}
 \opH^\mathrm{rsn}_{\res-\tmn} = g_{1a} a^\dag \sigma_1^{ge} + g_{1b} b^\dag \sigma_1^{ef} + g_{2b} b^\dag \sigma_2^{ge} + \mathhc,
\end{equation}
and the dispersive term $\textstyle \opH^\mathrm{disp}_\mathrm{r-a}$ reads as
\begin{equation} \label{eq:ham_disp}
	\opH^\mathrm{dsp}_{\res-\tmn} = \eta_{1a} a^\dag \sigma_1^{ef} + \eta_{1b} b^\dag \sigma_1^{ge} + \eta_{2b} b^\dag \sigma_2^{ef} + \mathhc.
\end{equation}
Parameters $\textstyle g_{1a}$ and $\textstyle g_{1b}$ are the coupling strengths of $\textstyle |g\rangle_1 \leftrightarrow |e\rangle_1$ and $\textstyle |e\rangle_1 \leftrightarrow |f\rangle_1$ transitions of the first transmon to resonators $\textstyle A$ and $\textstyle B$, respectively.
Parameter $\textstyle g_{2b}$ stands for the coupling strength of $\textstyle |g\rangle_2 \leftrightarrow |e\rangle_2$ transition of the second transmon to resonator $\textstyle B$.
Parameter $\textstyle \eta_{1a}$ is the coupling strength of resonator $\textstyle A$ to the first transmon $\textstyle |e\rangle_1 \leftrightarrow |f\rangle_1$ transition, $\textstyle \eta_{1b}$ stands for the coupling strength of resonator $\textstyle B$ to $\textstyle |g\rangle_1 \leftrightarrow |e\rangle_1$ transition of the first transmon, and $\textstyle \eta_{2b}$ denotes the coupling strength of resonator $\textstyle B$ to $\textstyle |e\rangle_2 \leftrightarrow |f\rangle_2$ transition of the second transmon.
For the transmon-type artificial atoms, one has $\textstyle \eta_{1a} \approx \sqrt{2}g_{1a}$, $\textstyle \eta_{1b}\approx g_{1b}/\sqrt{2}$, and $\textstyle \eta_{2b}\approx\sqrt{2}g_{2b}$ \cite{koch2007}.

The waveguides are modeled by the continua of non-interacting bosonic modes with the Hamiltonian $\textstyle \opH_\mathrm{w}$ given by
\begin{equation}
  \opH_\wg = \int^\infty_0 \dd\nu \nu \left(\alpha^\dag_\nu \alpha_\nu + \beta^\dag_\nu \beta_\nu\right),
\end{equation}
where $\textstyle \alpha_\nu (a^\dag_\nu)$ and $\textstyle \beta_\nu (\beta^\dag_\nu)$ stand for the annihilation (creation) operators of the photon with frequency $\nu$ propagating in the waveguide $\alpha$ and $\beta$, respectively.

The Hamiltonian $\opH_{\res-\wg}$, describing the coupling between the resonators and waveguides, has the form:
\begin{equation} \label{eq:ham_rw}
		 \opH_{\res-\wg} = \int^\infty_0 \dd \nu \left[f_a(\nu) a^\dag \alpha_\nu + f_b(\nu) b^\dag \beta_\nu + \mathhc\right],
\end{equation}
where $\textstyle f_a(\nu)$ and $\textstyle f_b(\nu)$ are the coupling strengths of resonator $\textstyle A$ to waveguide $\textstyle \alpha$ and resonator $\textstyle B$ to waveguide $\textstyle \beta$, correspondingly.

Note that the model Hamiltonian in Eq.~\eqref{eq:ham_1} does not explicitly include the Purcell filters.
Their contribution is absorbed into the frequency-dependent resonator-waveguide couplings $f_{a}(\nu)$ and $f_{b}(\nu)$.
The effect of the Purcell filters is outlined in Appendix~\ref{sec:purcell}.

\begin{figure*}[t!] 
	\centering
	\includegraphics{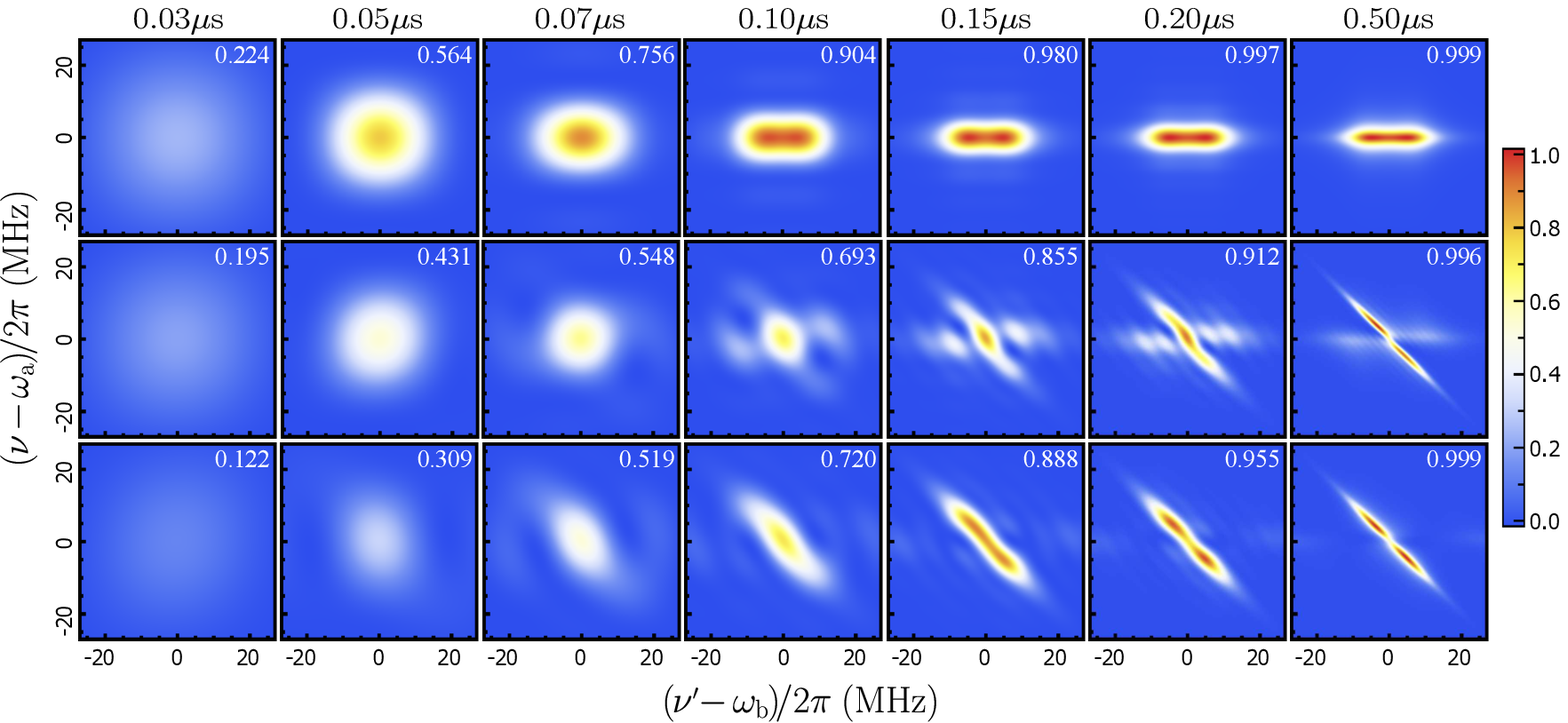}
	\caption{The snapshots of the joint spectra of the emitted photon pairs for the different moments of time.
		Parameters of the system are as follows: $\textstyle (\omega^{ge}_1 - \omega^{ef}_1)/2\pi = 400\,\mathrm{MHz}$, $\textstyle \kappa_a=\kappa_b =\kappa$, $\textstyle \kappa/2\pi = 25\,\mathrm{MHz}$, $\textstyle g_{1a}/2\pi = 5\,\mathrm{MHz}$, $\textstyle g_{2b}/2\pi = 10\,\mathrm{MHz}$, (upper row) $\textstyle g_{1b}=0$; (middle row) $\textstyle g_{1b}/2\pi = 10\,\mathrm{MHz}$; (bottom row) $\textstyle g_{1b}/2\pi = 25\,\mathrm{MHz}$.
		For aiding the visualization, each plot were normalized on the corresponding maximum value of the joint spectrum $\textstyle \max\{S_{\nu,\nu'}(t)\}$ and then multiplied on the probability $\textstyle p_{\alpha\beta}(t)$ of finding both waveguides containing photons at time $\textstyle t$.
		The corresponding values of $\textstyle p_{\alpha\beta}(t)$ are shown in each plot.
		\label{fig:fig_3}}
\end{figure*}

\subsection*{The effective Hamiltonian} \label{sec:effham}
The dispersive coupling term $\textstyle \opH^\mathrm{dsp}_\mathrm{r-a}$ given by Eq.~\eqref{eq:ham_disp} can be treated as a perturbation of the Hamiltonian $\textstyle \opH_\mathrm{sys}$ provided that $\textstyle |\lambda_{1a}|, |\lambda_{1b}|, |\lambda_{2b}| \ll 1$, where
\begin{equation*}
		\lambda_{1a} = \frac{\eta_{1a}}{\omega_1^{ef} - \omegaa}, \,
		\lambda_{1b} = \frac{\eta_{1b}}{\omega_1^{ge} - \omegab}, \,
		\lambda_{2b} = \frac{\eta_{2b}}{\omega_2^{ef} - \omegab}.
\end{equation*}
In this case, one can eliminate $\textstyle \opH^\mathrm{dsp}_\mathrm{r-a}$ via the Schrieffer-Wolff-type transformation $\textstyle \opH_\sys \rightarrow \opU^\dag \opH_\sys \opU$ with $\opU$ being the unitary operator given by \cite{klimov2000, blais2004}
\begin{equation} \label{eq:Udisp}
	\opU = \exp\left(\lambda_{1a} a^\dag \sigma_1^{ef} + \lambda_{1b} b^\dag \sigma_1^{ge} + \lambda_{2b} b^\dag \sigma_2^{ef} - \mathhc\right).
\end{equation}
Keeping the terms contributing up to the first order in the small parameters $\textstyle \lambda_{1a}$, $\textstyle \lambda_{1b}$, and $\textstyle \lambda_{2b}$, one arrives at the effective Hamiltonian $\textstyle \Heff$ of the form (see details of derivation in Appendix~\ref{sec:appb}):
\begin{equation} \label{eq:ham_eff}
 \Heff = \opbH_\tmn + \opbH_\res + \opH^\mathrm{rsn}_{\res-\tmn} + \opH_\mathrm{rr-tm} + \opH_\wg + \opH_{\res-\wg}.
\end{equation}
In the dressed basis, the Hamiltonian describing the transmons reads as
\begin{equation} \label{eq:ham2_a}
 \begin{split}
 	  \opbH_\tmn = \bar{\omega}_1^{ge} \sigma_1^{ee} + \bar{\omega}_1^{gf} \sigma_1^{ff}
 	  + \omega_2^{ge} \sigma_1^{ee} + \bar{\omega}_2^{gf} \sigma_2^{ff},
 \end{split}
\end{equation}
with $\textstyle \bar{\omega}_1^{ge} = \omega_1^{ge} + \chi_{1b}$, $\textstyle \bar{\omega}_1^{gf} = \bar{\omega}_1^{ge} + \bar{\omega}_1^{ef} = \omega_1^{ef} + \chi_{1a}$, and $\textstyle \bar{\omega}_2^{gf} = \omega_2^{ge} + \bar{\omega}_2^{ef} = \omega_2^{gf} + \chi_{2b}$ being the dressed (renormalized) frequencies of $\textstyle |g\rangle_1 \leftrightarrow |e\rangle_1$ and $\textstyle |g\rangle_1 \leftrightarrow |f\rangle_1$ transitions of the first transmon, and $\textstyle |g\rangle_2 \leftrightarrow |f\rangle_2$ transition of the second transmon, respectively.
Here, we introduced the notations $\textstyle \chi_{1a} = \lambda_{1a}\eta_{1a}$, $\textstyle \chi_{1b} = \lambda_{1b}\eta_{1b}$, and $\textstyle \chi_{2b} = \lambda_{2b}\eta_{2b}$.

The Hamiltonian of the resonators acquires the form
\begin{equation}
	\begin{split}
	 \opbH_\res = & \, \big(\omegaa + \chi_{1a} \opZ_1^{fe}\big) a^\dag a \\
	 & \, + \big(\omegab + \chi_{1b} \opZ_1^{eg} + \chi_{2b} \opZ_2^{fe}\big) b^\dag b,
	\end{split}
\end{equation}
where we introduced the notation $\textstyle \opZ_j^{ss'} \equiv \sigma_j^{ss} - \sigma_j^{s's'}$.

The term $\textstyle \opH_{\res\res-\tmn}$ in Eq.~\eqref{eq:ham_eff} is given by
\begin{equation} \label{eq:ham_rrq}
 \opH_{\res\res-\tmn} = \varUpsilon \big(a^\dag b^\dag \sigma_1^{gf} + \sigma_1^{fg} a \, b\big),
\end{equation}
where $\textstyle 2\varUpsilon = \lambda_{1a}\eta_{1b} - \lambda_{1b}\eta_{1a}$.
This term describes the processes of simultaneous exchange of two excitations between the first transmon and resonators $\textstyle A$ and $\textstyle B$.

In the further analysis, we work in the dressed basis and describe the dynamics of the system using the Hamiltonian $\textstyle \Heff$ expressed by Eqs.~\eqref{eq:ham_eff}--\eqref{eq:ham_rrq}.

Note that here we focus on the unitary dynamics of the system, neglecting the dissipation.
We assume that dissipation processes occur on the timescales much longer than the timescales of the coherent processes in the system.
With such an approach, we simplify the theoretical treatment of the considered system while still grasping the essential features of its quantum dynamics.

Indeed, for the system parameters we work with, the photon leakage rates from the resonators to the output waveguides are $\textstyle \kappa/2\pi>1\,\mathrm{MHz}$, while the internal quality factors of the CPW resonators can exceed $\textstyle 10^6$ \cite{megr2012, bruno2015}, giving the intrinsic resonator photon loss rates $\textstyle \gamma_\mathrm{res}/2\pi \lesssim 0.01\,\mathrm{MHz}$.
The coaxial resonators \cite{reagor2016, heidler2021} and the three-dimensional microwave cavities \cite{paik2011, reagor2013, flurin2015} offer even higher internal quality factors reaching up to $10^8-10^9$.
Thus, the process of photon leakage to the output waveguides dominates over the photon dissipation inside the resonators.
The typical excited-state lifetimes of the state-of-the-art transmons reach $\textstyle 0.05-0.1\,\mathrm{ms}$ \cite{barends2013, riste2017}, with the recent experiments \cite{place2021, wang2022} reporting lifetimes up to $\textstyle 0.5\,\mathrm{ms}$, which are more than an order of magnitude longer than the times the transmons dwell in their excited states for the system parameters we consider in this work. 
Thus, the decoherence of transmons has a negligible effect on the emission dynamics.

\section{Emission dynamics} \label{sec:evol}

We assume that initially (at $\textstyle t=0$), the system is prepared in the state $\textstyle |\Psi_\mathrm{in}\rangle$ with the transmons residing in their excited states $\textstyle |e\rangle_1$ and $\textstyle |e\rangle_2$, while the resonators and the waveguides are void of excitation.
Thus, the initial state of the system reads as $\textstyle |\Psi_\mathrm{in}\rangle = \sigma_1^{eg}\sigma_2^{eg}\vacket$,
where $\textstyle \vacket = |\varnothing\rangle_\alpha|\varnothing\rangle_\beta |0\rangle_a |0\rangle_b |g\rangle_1 |g\rangle_2$ stands for the vacuum state of the system -- a state with a void of photons in the resonators and the waveguides and the transmons residing in their ground states.
In our analysis, we do not account for the effect of thermal excitations on the dynamics of the system.
This simplification is justified since the superconducting circuit QED systems usually operate at temperatures $\textstyle T_\sys \sim 10-30\,\mathrm{mK}$, while the typical working frequencies of microwave circuit QED setups are $\omega_\sys/2\pi \sim 5-20\,\mathrm{GHz}$ \cite{gu2017, krantz2019}.
For these parameters, one obtains the estimate $\textstyle n_\mathrm{th}<10^{-3}$ for the average number of thermal excitations in the system.

At the arbitrary moment of time $\textstyle t$, the state of the system $\textstyle |\varPsi(t)\rangle = \ee^{-\ii \Heff t}|\Psi_\mathrm{in}\rangle$ is expressed as
\begin{widetext}
\begin{equation} \label{eq:wavefunc}
	\begin{split}
	 |\varPsi(t)\rangle = & \, \int^\infty_0 \dd\nu \int^\infty_0 \dd\nu' \varPhi_{\nu,\nu'} (t) \alpha^\dag_{\nu} \beta^\dag_{\nu'}\vacket + \int^\infty_0 \dd\nu \left[\varXi^{\alpha}_\nu(t) \alpha^\dag_{\nu} b^\dag + \varXi^{\beta}_\nu(t) \beta^\dag_{\nu}a^\dag + \varTheta^{\alpha}_\nu(t) \alpha^\dag_{\nu}\sigma_2^{eg} + \varTheta^{\beta}_\nu(t)\beta^\dag_{\nu}\sigma_1^{eg}\right]\vacket \\
	 & \, + R(t) a^\dag b^\dag\vacket + Q(t) \sigma_1^{fg}\vacket + Y_{a}(t) a^\dag \sigma_2^{eg}\vacket + Y_{b}(t) b^\dag \sigma_1^{eg}\vacket + X(t)\sigma_1^{eg}\sigma_2^{eg} \vacket.
	\end{split}
\end{equation}
\end{widetext}
The first term in Eq.~\eqref{eq:wavefunc} corresponds to the state of the system with both waveguides hosting a photon, with $\textstyle \varPhi_{\nu,\nu'}$ being the joint spectral amplitude.
The quantity $\textstyle S_{\nu,\nu}(t) = |\varPhi_{\nu,\nu'}(t)|^2$ determines the joint spectrum -- the probability density distribution of finding a photon with frequency $\textstyle \nu$ propagating in waveguide $\textstyle \alpha$ and a photon with frequency $\textstyle \nu'$ propagating in waveguide $\textstyle \beta$ at the moment of time $\textstyle t$.
The probability of finding the waveguides both accommodating photons is given by
\begin{equation} \label{eq:prob_2w}
	p_{\alpha\beta}(t) = \int^\infty_0 \dd\nu \int^\infty_0 \dd\nu' \, S_{\nu,\nu'}(t). 
\end{equation}

The remaining terms in the upper line of Eq.~\eqref{eq:wavefunc} correspond to the states with photon propagating in only one of the waveguides: $\textstyle \varXi_\nu^{\alpha} (\varXi_\nu^{\beta})$ is the amplitude of the state with the photon in resonator $\textstyle A (B)$ and the transmons in their ground states, $\textstyle \varTheta_\nu^{\alpha} (\varTheta_\nu^{\beta})$ is the amplitude of the state with the resonators void of photons and the first (second) transmon residing in the excited state $\textstyle |e\rangle_{1 (2)}$.
The terms in the bottom line of Eq.~\eqref{eq:wavefunc} correspond to the states of the system with both waveguides void of excitations.

The equations of motion governing the evolution of the probability amplitudes in Eq.~\eqref{eq:wavefunc} can be compactly written as
\begin{equation} \label{eq:eqmot}
	\frac{\partial\mathbf{\boldsymbol{\mu}}(t)}{\partial t} = - \ii \, \boldsymbol{\Omega} \, \boldsymbol{\mu}(t),
\end{equation}
\begin{widetext}
where $\textstyle \boldsymbol{\mu}(t) = [\varPhi_{\nu,\nu'}(t), \varXi^{\alpha}_\nu(t), \varXi^{\beta}_\nu(t), \varTheta^{\alpha}_\nu(t), \varTheta^{\beta}_\nu(t), R(t), Q(t), Y_a(t), Y_b(t), X(t)]^\mathrm{T}$ and the matrix $\textstyle \boldsymbol{\Omega}$ reads as
\begin{equation*} \label{eq:mtrx}
	\boldsymbol{\Omega} =
    \begin{pmatrix}
	 \nu+\nu' 	& f_b(\omega_b)		& f_a(\omega_a) & 0 	 & 0      & 0   & 0   & 0 & 0 & 0 \\
	 0 			& \nu + \widetilde{\omega}_b - \chi_{1b} & 0 					 & g_{2b} & 0      & f_a(\omega_a) & 0   & 0 & 0 & 0 \\
	 0 			& 0								& \nu' + \widetilde{\omega}_a 		 & 0      & g_{1a} & f_b(\omega_b) & 0   & 0 & 0 & 0 \\
	 0 			& g_{2b} 						& 0	& \nu + \omega_2^{ge} & 0 	 & 0 	  & 0 	& f_a(\omega_a) & 0 & 0 \\
     0 & 0 & g_{1a} & 0 & \nu' + \bar{\omega}_1^{ge} & 0 & 0 & 0 & f_b(\omega_b) & 0 \\
     0 & 0 & 0 & 0 & 0 & \widetilde{\omega}_a + \widetilde{\omega}_b-\chi_{1b} & \varUpsilon & g_{2b} & g_{1a} & 0 \\
	 0 & 0 & 0 & 0 & 0 & \varUpsilon & \bar{\omega}^{gf}_1 & 0 & g_{1b} & 0 \\
     0 & 0 & 0 & 0 & 0 & g_{2b} & 0 & \widetilde{\omega}_a + \omega_2^{ge} & 0 & g_{1a} \\
     0 & 0 & 0 & 0 & 0 & g_{1a} & g_{1b} & 0 & \widetilde{\omega}_b + \bar{\omega}_1^{ge} + \chi_{1b} & g_{2b} \\ 	 
     0 & 0 & 0 & 0 & 0 & 0 & 0 & g_{1a} & g_{2b} & \bar{\omega}_1^{ge} + \omega_2^{ge} \\
    \end{pmatrix},     
\end{equation*}
\end{widetext}
where $\textstyle \widetilde{\omega}_a = \omegaa - \ii \kappa_{a}/2$ and $\textstyle \widetilde{\omega}_b = \omegab - \ii \kappa_{b}/2$ with
$\textstyle \kappa_{a} = 2\pi f^2_{a}(\omegaa)$ and $\textstyle \kappa_{b} = 2\pi f^2_{b}(\omegab)$ being the photon leakage rate from resonators $\textstyle A$ and $\textstyle B$ into waveguides $\textstyle \alpha$ and $\textstyle \beta$, correspondingly.
The derivation of Eq.~\eqref{eq:eqmot} is given in Appendix~\ref{sec:appd} and follows the approach demonstrated in Refs.~\cite{sto2019} and \cite{sok2020}.

For computing the probability amplitudes standing in Eq.~\eqref{eq:wavefunc}, we use the symbolic solution of Eq.~\eqref{eq:eqmot} expressed as $\textstyle \boldsymbol{\mu}(t) = \exp(-\ii \boldsymbol{\Omega} t) \boldsymbol{\mu}(0)$.
The matrix exponential is evaluated numerically using the built-in function \texttt{MatrixExp} of \textsc{Mathematica}.

Figure~\ref{fig:fig_3} provides the series of snapshots of the joint spectra of the emitted photons evaluated for the specific moments of time and various parameters of the system.
The computations reveal that switching on the coupling between resonator $\textstyle B$ and $\textstyle |e\rangle_1 \leftrightarrow |f\rangle_1$ transition of the first transmon results in the pronounced frequency anti-correlation of the emitted photons, which manifests itself as an accumulation of the joint spectrum near the line given by $\textstyle \nu + \nu' = \omegaa + \omegab$.

\begin{figure}[t!]
	\centering
	\includegraphics{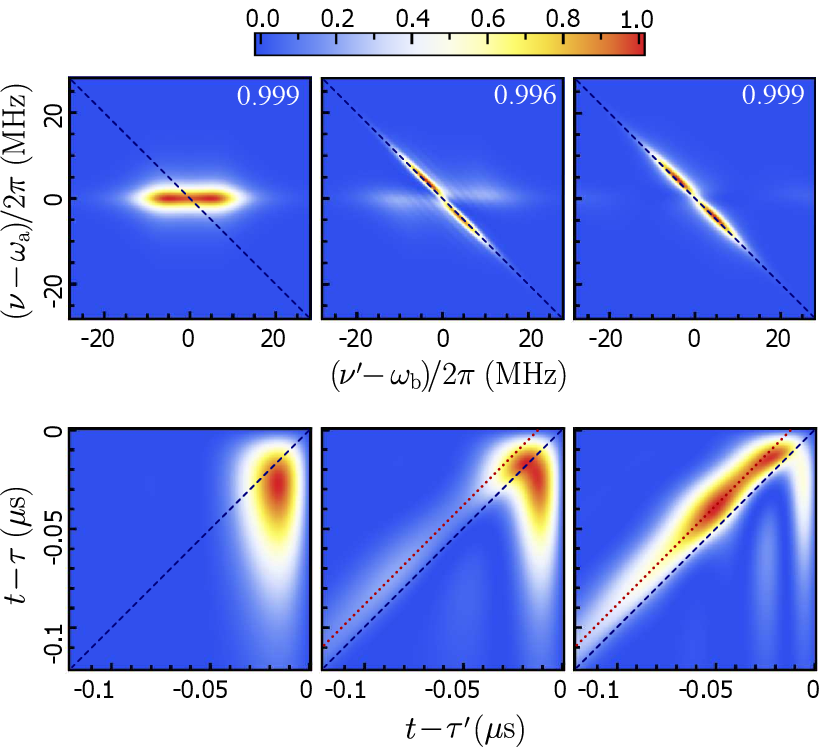}
	\caption{(Upper row) Joint spectra of the emitted photons $\textstyle S_{\nu,\nu'}(t)$ versus (bottom row) the corresponding time-domain probability densities $\textstyle A_{\tau,\tau'}(t)$ at  $\textstyle t=0.5\,\mathrm{\mu s}$ for different values of $\textstyle g_{1b}$: (left column) $\textstyle g_{1b} = 0$; (central column) $\textstyle g_{1b}/2\pi = 10\,\mathrm{MHz}$; (right column) $\textstyle g_{1b}/2\pi = 25\,\mathrm{MHz}$.
	The rest of the parameters are as described in the caption of Fig.~\ref{fig:fig_3}.
	The dashed lines correspond $\textstyle \nu+\nu'=\omega_a + \omega_b$ (upper row) and $\textstyle \tau' = \tau$ (bottom row).
	Dotted lines are defined as $\textstyle \tau' = \tau + \delta\tau$.
	Plots were normalized on $\textstyle \max\{S_{\nu,\nu'}(t)\}$ (upper row) and $\textstyle \max\{A_{\tau,\tau'}(t)\}$ (bottom row) and then multiplied on the corresponding value of $\textstyle p_{\alpha\beta}(t)$. \label{fig:fig_4}}
\end{figure}

For the analysis of the temporal properties of the emitted photons, we introduce their time-domain joint probability amplitude $\textstyle \phi_{\tau,\tau'}(t)$ expressed via the Fourier transform of the joint spectral amplitude $\textstyle \varPhi_{\nu,\nu'}(t)$ as:
\begin{equation}
	\phi_{\tau,\tau'}(t) = \frac{1}{2\pi}\int\dd\nu \, \ee^{\ii\nu \tau} \int \dd\nu' \, \ee^{\ii \nu' \tau'} \, \varPhi_{\nu,\nu'}(t).
\end{equation}
One can interpret the quantity $\textstyle A_{\tau,\tau'}(t)=|\phi_{\tau,\tau'}(t)|^2$ as the distribution of the joint probability density of finding one photon at point $\textstyle x=\upsilon_\alpha t$ in waveguide $\textstyle \alpha$ at instant $\textstyle \tau$ along with finding another photon at point $\textstyle x=\upsilon_\beta t$ in waveguide $\textstyle \beta$ at instant $\textstyle \tau'$, assuming that the coupling points of the waveguides to the resonators are at $\textstyle x=0$.
Here, $\textstyle \upsilon_\alpha$ and $\textstyle \upsilon_\beta$ stand for the photon group velocity in waveguide $\textstyle \alpha$ and $\textstyle \beta$, correspondingly.
Figure~\ref{fig:fig_4} demonstrates the joint spectra $\textstyle S_{\nu,\nu'}(t)$ of the emitted photons at $\textstyle t=0.5\,\mathrm{\mu s}$ versus the respective time-domain probability densities $\textstyle A_{\tau,\tau'}(t)$  for different values of $\textstyle g_{1b}$.
One can notice that the frequency anti-correlation of the emitted photons arising for $\textstyle g_{1b}\neq 0$ is accompanied by their time correlation, which appears as a dilution of the time-domain probability density for $\textstyle \tau'\leq\tau$ and its aggregation in the region $\textstyle \tau' > \tau$ along the line $\textstyle \tau'=\tau + \delta\tau$.
This result indicates that when we switch on the interaction between the second transmon and resonator $\textstyle B$, the emission of the photon into waveguide $\textstyle \beta$ starts \emph{after} the emission of the photon into waveguide $\textstyle \alpha$.
The delay between the photons is roughly estimated as $\textstyle \delta\tau \approx (\kappa_b/2)^{-1}$.
Such behavior is consistent with the considerations presented in Sec.~\ref{sec:scheme}.

Thus, the general idea discussed in Sec.~\ref{sec:scheme} is supported by the numerical results confirming that the proposed setup can emit the TF entangled photon pairs.
As a quantitative measure of photon entanglement, we use the von Neumann entropy, whose definition, along with the extensive analysis of its dependence of the parameters of the system, is given in Sec.~\ref{sec:entang}.

\begin{figure*}[t!] 
	\centering
	\includegraphics{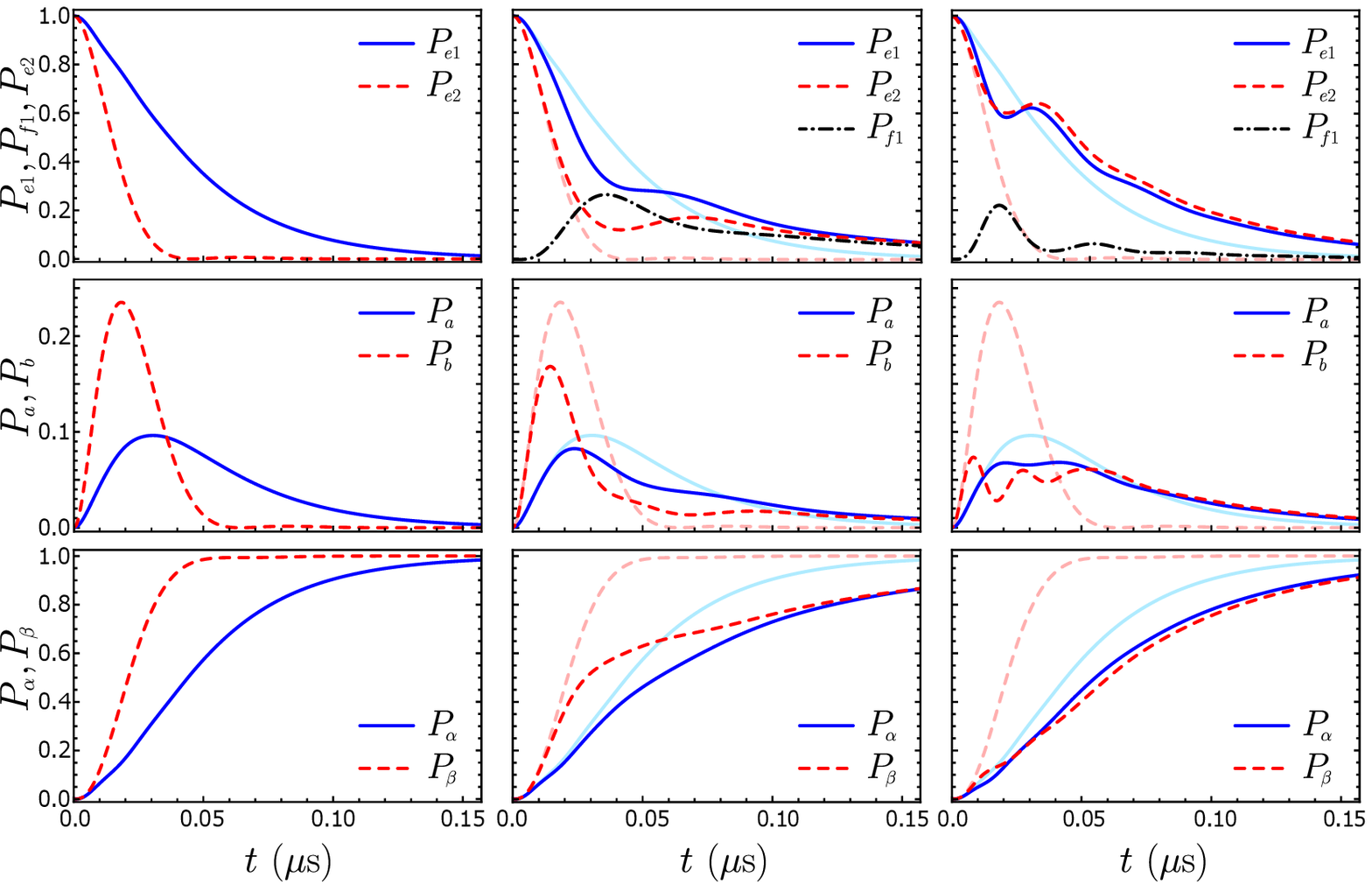}
	\caption{Dynamics of the populations of the transmons eigenlevels (upper row), resonators populations (middle row), and waveguides populations (bottom row). Parameters of the system used for computations are the following: (left column) $\textstyle g_{1b}=0$; (middle column) $\textstyle g_{1b}/2\pi = 10\,\mathrm{MHz}$; (right column) $\textstyle g_{1b}/2\pi = 25\,\mathrm{MHz}$.
		The rest of the parameters are the same as in Fig.~\ref{fig:fig_4}.
		For a better perception of the numerical results, we supplement the plots for $\textstyle g_{1b}\neq 0$ (middle and right columns) with the corresponding curves (shown in lower saturation) for the decoupled case $\textstyle g_{1b} = 0$.
		\label{fig:fig_5}}
\end{figure*}

Using the time-dependent wavefunction of the system, one can compute various observables such as populations of the transmon eigenlevels, the resonators, and the waveguides at arbitrary moments of time.
The population of $|s\rangle_j$ eigenlevel of the $\textstyle j$-th transmon is given by $\textstyle P_{sj}(t) = \langle \varPsi(t)|\sigma^{ss}_{j}|\varPsi(t)\rangle$.
With the wavefunction expressed by Eq.~\eqref{eq:wavefunc}, one obtains
\begin{equation} \label{eq:pop_1}
	\begin{split}
	 P_{e1}(t) & \, = |X(t)|^2 + |Y_b(t)|^2 + \int^\infty_0 \dd\nu |\varTheta^\beta_\nu(t)|^2, \\ P_{f1}(t) & \, = |Q(t)|^2
	\end{split}
\end{equation}
for the populations of $|e\rangle_1$ and $|f\rangle_1$ eigenlevels of the first transmon, respectively.
The population of $\textstyle |e\rangle_2$ eigenlevel of the second transmon is given by
\begin{equation}
	P_{e2}(t) = |X(t)|^2 + |Y_a(t)|^2 + \int^\infty_0 \dd\nu |\varTheta^\alpha_\nu(t)|^2.
\end{equation}
The populations of resonators $\textstyle A$ and $\textstyle B$ are expressed in terms of the probability amplitudes as
\begin{subequations}
 \begin{equation}
 	\begin{split}
 	  P_a(t) & \, = \langle\varPsi(t)|a^\dag a|\varPsi(t)\rangle \\
 	         & \, = |R(t)|^2 + |Y_a(t)|^2 + \int^\infty_0 \dd\nu |\varXi^\beta_\nu(t)|^2,
 	\end{split}
 \end{equation}

 \begin{equation}
 	\begin{split}
   	 P_b(t) = & \, \langle\varPsi(t)|b^\dag b|\varPsi(t)\rangle \\
	          & \, = |R(t)|^2 + |Y_b(t)|^2 + \int^\infty_0 \dd\nu |\varXi^\alpha_\nu(t)|^2.
	\end{split}
 \end{equation}
\end{subequations}
Finally, the waveguide populations are computed as
\begin{subequations}
 \begin{equation}
 	\begin{split}
 	  P_\alpha(t) & \, = \int^\infty_0 \dd\nu \, \langle\varPsi(t)|\alpha^\dag_\nu \alpha_\nu|\varPsi(t)\rangle \\
 		          & \, = p_{\alpha\beta}(t) + \int^\infty_0 \dd\nu \left[|\varXi^\alpha_\nu(t)|^2 + |\varTheta^\alpha_\nu(t)|^2\right],
 	\end{split}
 \end{equation}
 \begin{equation}
  	\begin{split}
  		P_\beta(t) & \, = \int^\infty_0 \dd\nu \, \langle\varPsi(t)|\beta^\dag_\nu \beta_\nu|\varPsi(t)\rangle \\
  		           & \, = p_{\alpha\beta}(t) + \int^\infty_0 \dd\nu \left[|\varXi^\beta_\nu(t)|^2 + |\varTheta^\beta_\nu(t)|^2\right],
  	\end{split}
  \end{equation}
\end{subequations}
where $\textstyle p_{\alpha\beta}(t)$ is given by Eq.~\eqref{eq:prob_2w}.

Figure~\ref{fig:fig_5} demonstrates the dynamics of the populations of the transmons eigenlevels as well as the populations of the resonators and the waveguides for different values of coupling $\textstyle g_{1b}$ between resonator $\textstyle B$ and $\textstyle |e\rangle_1 \leftrightarrow |f\rangle_1$ transition of the first transmon.
In agreement with the earlier qualitative considerations and numerical results, we observe that switching on the coupling between the first transmon and resonator $\textstyle B$ leads to pronounced modification of the emission dynamics compared to the decoupled case.

\begin{figure*}[t!] 
	\centering
	\includegraphics{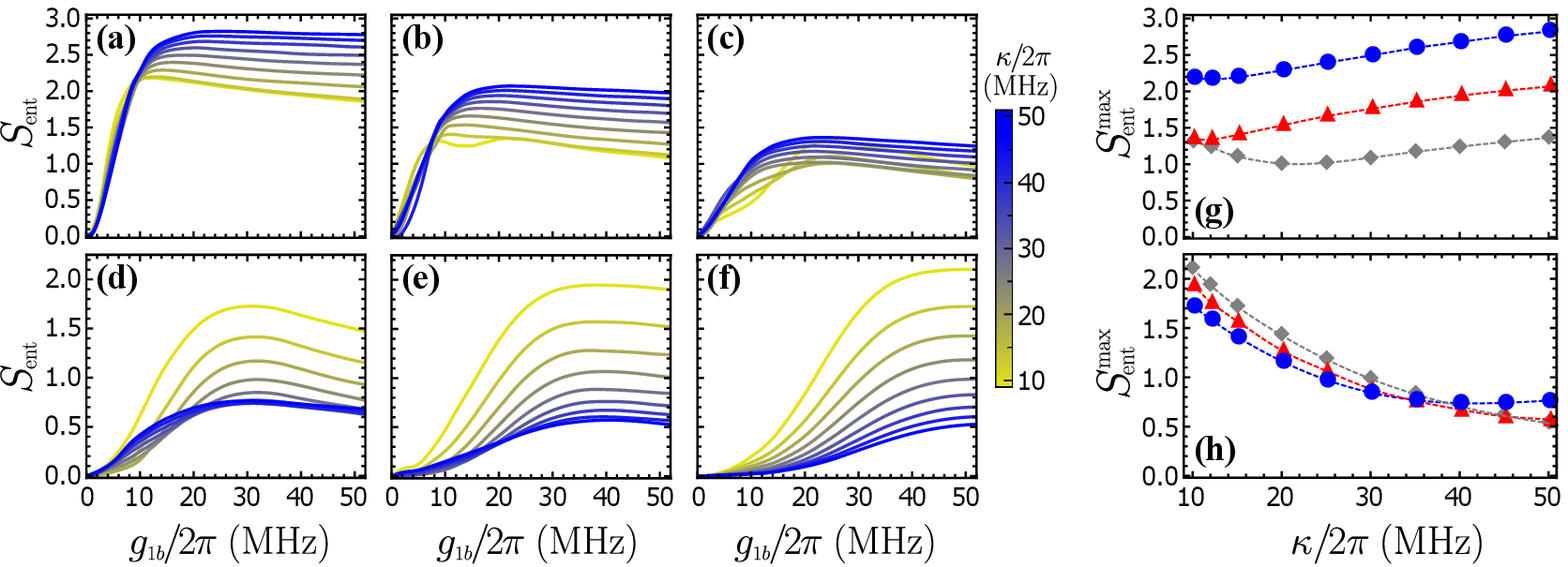}
	\caption{Plots (a)--(f) show the dependence of the entanglement entropy $\textstyle S_\mathrm{ent}$ on $\textstyle g_{1b}$ for different values of $\textstyle g_{1a}$ and $\textstyle \kappa$ (encoded by the color gradient), where we set $\textstyle (\kappa_a = \kappa_b) = \kappa$.
		Parameters of the system used for computations are the following: (a) $\textstyle g_{1a}/2\pi = 5\,\textrm{MHz}$, (b) $\textstyle g_{1a}/2\pi = 7\,\textrm{MHz}$, (c) $\textstyle g_{1a}/2\pi = 10\,\textrm{MHz}$, (d) $\textstyle g_{1a}/2\pi = 15\,\textrm{MHz}$, (e) $\textstyle g_{1a}/2\pi = 20\,\textrm{MHz}$, (f) $\textstyle g_{1a}/2\pi = 30\,\textrm{MHz}$.
		Plots (g) and (h) demonstrate the effect of $\textstyle \kappa$ on the maximal value of the entanglement entropy $\textstyle S^\mathrm{max}_\mathrm{ent}$, which can be achieved for the given values of $\textstyle g_{1a}$ and $\textstyle g_{2b}$.
		Parameters used in plot (g): $\textstyle g_{1a}/2\pi = 5\,\mathrm{MHz}$ (blue circles),  $\textstyle g_{1a}/2\pi = 7\,\mathrm{MHz}$ (red triangles), $\textstyle g_{1a}/2\pi = 10 \,\mathrm{MHz}$ (grey diamonds).
		Parameters used in plot (h): $\textstyle g_{1a}/2\pi = 15\,\mathrm{MHz}$ (blue circles), $\textstyle g_{1a}/2\pi = 20\,\mathrm{MHz}$ (red triangles), $\textstyle g_{1a}/2\pi = 30\,\mathrm{MHz}$ (grey diamonds).
		For all plots, we set $\textstyle g_{2b}/2\pi = 10\,\textrm{MHz}$.
		\label{fig:fig_6}}
\end{figure*}

\section{Photon entanglement} \label{sec:entang}

In the course of its evolution, the considered system eventually reaches the state when all emission processes are finished, the transmons have relaxed to their ground states, and the photons propagate in the waveguides as free excitations.
This final state of the system $\textstyle |\Psi_\mathrm{fi}\rangle$ is given by
\begin{equation} \label{eq:psi_out}
	|\Psi_\mathrm{fi}\rangle = \int \dd\nu \int \dd\nu' \Phi^\mathrm{out}_{\nu, \nu'} \alpha^\dag_\nu \beta^\dag_{\nu'}\vacket,
\end{equation}
with $\textstyle \Phi^\mathrm{out}_{\nu, \nu'}$ being the joint spectral amplitude of the outgoing photon pair.
If the emitted photons are not entangled, their joint spectral amplitude can be factorized into a product of single-photon amplitudes $\textstyle \Phi^\mathrm{out}_{\nu, \nu'} = \phi_\alpha(\nu)\phi_\beta(\nu')$, while the joint spectral amplitude of TF entangled photon pair is not factorable $\textstyle \Phi^\mathrm{out}_{\nu,\nu'}\neq\phi_\alpha(\nu)\phi_\beta(\nu')$.
For testing the factorizability of the joint spectral amplitude of the emitted photons $\textstyle \Phi^\mathrm{out}_{\nu, \nu'}$, we perform its Schmidt decomposition \cite{law2000, eberly2006}:
\begin{equation} \label{eq:decomp}
	\Phi^\mathrm{out}_{\nu, \nu'} = \sum_{j} \sqrt{\lambda_j} \, \varphi_{j,\nu} \vartheta_{j,\nu'},
\end{equation}
where the weights $\textstyle \lambda_j\geq 0$ satisfying the condition $\textstyle \sum_j \lambda_j = 1$ are usually referred to as the Schmidt coefficients, and the single-photon spectral amplitudes $\textstyle \varphi_{j,\nu}$ and $\textstyle \vartheta_{j,\nu}$ are called the Schmidt modes.
The latter constitute a complete set of orthonormal functions obeying the relations
$\textstyle \sum_{j} \zeta^*_{j,\nu} \zeta_{j,\nu'} = \delta(\nu - \nu')$ and $\textstyle \int \dd \nu \, \zeta^*_{j,\nu} \zeta_{k,\nu} = \delta_{j,k}$,
where $\textstyle \zeta = \varphi$ or $\vartheta$.

The Schmidt coefficients $\textstyle \lambda_j$ are determined by solution of the eigenvalue problem \cite{law2000}:
\begin{equation} \label{eq:eigen}
	\begin{split}
		\int \dd \nu' \, \mathcal{K}^\varphi_{\nu, \nu'} \, \varphi_{j,\nu'} = \lambda_j \varphi_{j,\nu}, \\
		\int \dd \nu' \, \mathcal{K}^\vartheta_{\nu, \nu'} \, \vartheta_{j,\nu'} = \lambda_j \vartheta_{j,\nu},
	\end{split}
\end{equation}
where the integral kernels $\textstyle \mathcal{K}^{\varphi}_{\nu, \nu'}$ and $\textstyle \mathcal{K}^{\vartheta}_{\nu, \nu'}$ are defined as
\begin{equation} \label{eq:K_def}
	\begin{split}
		\mathcal{K}^{\varphi}_{\nu, \nu'} = \int \dd\varpi \, (\Phi^\mathrm{out}_{\nu,\varpi})^* \Phi^\mathrm{out}_{\nu',\varpi}, \\
		\mathcal{K}^{\vartheta}_{\nu, \nu'} = \int \dd\varpi \, (\Phi^\mathrm{out}_{\varpi,\nu})^* \Phi^\mathrm{out}_{\varpi,\nu'}.
	\end{split}
\end{equation}

\begin{figure*}[t!] 
	\centering
	\includegraphics{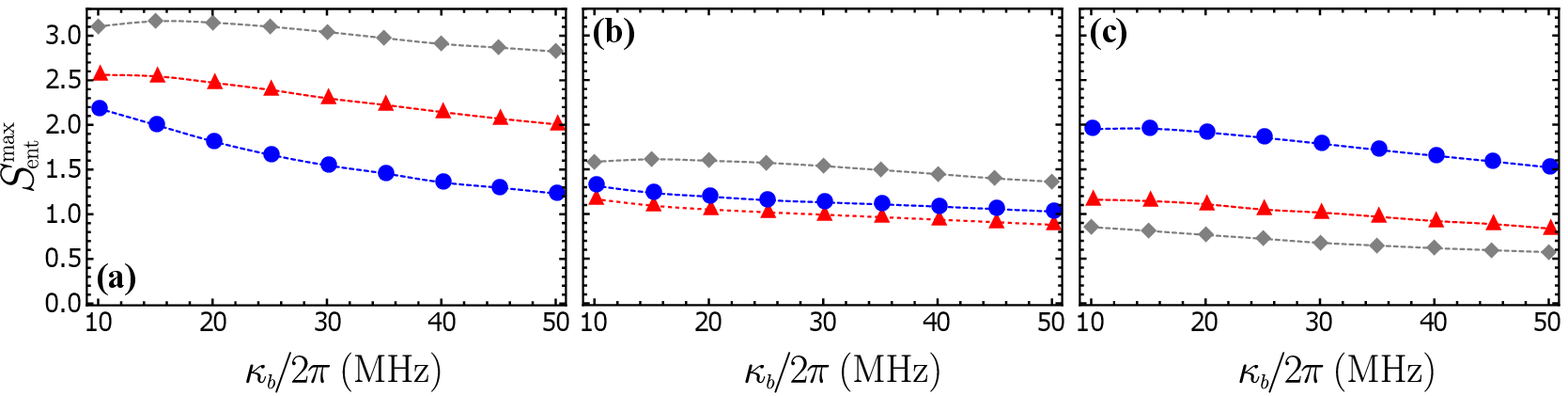}
	\caption{Dependence of $\textstyle S^\mathrm{max}_\mathrm{ent}$ on $\textstyle \kappa_b$ for different values of $\textstyle \kappa_a$: $\textstyle \kappa_a/2\pi = 10\,\mathrm{MHz}$ (blue circles), $\textstyle \kappa_a/2\pi = 25\,\mathrm{MHz}$ (red triangles), and $\textstyle \kappa_a/2\pi = 50\,\mathrm{MHz}$ (grey diamonds), and $\textstyle g_{1a}$: (a) $\textstyle g_{1a}/2\pi = 5\,\mathrm{MHz}$, (b) $\textstyle g_{1a}/2\pi = 10\,\mathrm{MHz}$, (c) $\textstyle g_{1a}/2\pi = 20\,\mathrm{MHz}$.
	The rest of the system parameters are the same as in Fig.~\ref{fig:fig_6}.
	\label{fig:fig_7}}
\end{figure*}

The eigenvalue problem in Eq.~\eqref{eq:eigen} is treated numerically using the tools provided by the \textsc{Mathematica} system.
We transform the eigenvalue problem for the integral operators into that for the matrices. 
For this purpose we discretize the kernels $\textstyle \mathcal{K}^{\varphi}_{\nu,\nu'}$ and $\textstyle \mathcal{K}^{\vartheta}_{\nu,\nu'}$ into a uniform $\textstyle \mathsf{N} \times \mathsf{N}$ grid on a square domain spanning $\textstyle \pm \varDelta\omega$ around the resonator frequencies $\textstyle \omega_a$ and $\textstyle \omega_b$, where $\textstyle \varDelta\omega$  satisfies the condition $\textstyle \int^{\omega_a+\varDelta\omega}_{\omega_a-\varDelta\omega} \dd\nu \int^{\omega_b+\varDelta\omega}_{\omega_b-\varDelta\omega} \dd\nu' |\Phi^\mathrm{out}_{\nu,\nu'}|^2 \geq \Theta$ with $\textstyle 0 < \Theta < 1$.
The values of $\textstyle \mathcal{K}^{\varphi}_{\nu,\nu'}$ and $\textstyle \mathcal{K}^{\vartheta}_{\nu,\nu'}$ in the grid nodes are computed by the numerical evaluation of the integrals in Eq.~\eqref{eq:K_def} using the \texttt{NIntegrate} function.
The joint spectral amplitude of the outgoing photons is evaluated as $\textstyle \Phi^\mathrm{out}_{\nu,\nu'} = \varPhi_{\nu,\nu'}(t_\infty)$, where the moment of time $\textstyle t_\infty$ is determined as $\textstyle p_{\alpha\beta}(t_\infty) > 0.999$.
Then, we use the \texttt{Eigenvalues} function for determining the set of eigenvalues $\textstyle \{\Lambda_j\}$ for the obtained matrices.
A finite discretization of the bounded domain of photon frequencies results in $\textstyle (\mathcal{I} \equiv \sum^\mathsf{N}_{j=1} \Lambda_j) < 1$, so we make a normalization $\textstyle \{\lambda_j\} = \{\Lambda_j\}/\mathcal{I}$ to ensure that $\textstyle \sum_{j} \lambda_j = 1$.
We set $\textstyle \mathsf{N}=100$ and $\textstyle \Theta = 0.99$ for all computations.
For the system parameters we use for computations, the extension of the frequency domain (by setting larger $\textstyle \Theta$) and using the finer grid (by increasing $\textstyle \mathsf{N}$) has only a minor effect on the evaluated values of the Schmidt coefficients and the entanglement entropies.

As a measure of entanglement of the emitted photons, we use the entanglement (von Neumann) entropy $\textstyle S_\mathrm{ent}$, which is expressed via the Schmidt coefficients as \cite{bennet1996}:
\begin{equation} \label{eq:entr_def}
	S_\mathrm{ent} = - \sum_j \lambda_j \log_2 \lambda_j.
\end{equation}
The non-zero entanglement entropy, $\textstyle S_\mathrm{ent} > 0$, implies that emitted photons are entangled.

Figure~\ref{fig:fig_6} aggregates the results of computations demonstrating the dependence of the entanglement entropy $\textstyle S_\mathrm{ent}$ of the emitted photons on the interrelation between the transmon-resonator coupling parameters and the photon leakage rate $\textstyle \kappa$ from the resonators to the corresponding waveguides for the case $\textstyle \kappa = (\kappa_a = \kappa_b)$.
Computations reveal that for $\textstyle g_{1a}<g_{2b}$, the entanglement entropy rapidly grows with the increase of $\textstyle g_{1b}$, reaching its maximum $\textstyle S^\mathrm{max}_\mathrm{ent}$ for some value of $\textstyle g_{1b}$, then slowly decreasing with the further increase of $\textstyle g_{1b}$.
In this regime, faster photon leakage to the waveguides (i.e., shorter photon lifetimes inside the resonators) leads to stronger photon entanglement, as illustrated in Figs.~\hyperref[fig:fig_6]{\ref*{fig:fig_6}(g)}.
The increase of the ratio $\textstyle g_{1a}/g_{2b}$ eventually leads to the opposite behavior, when the entanglement weakens with the increase of the photon leakage rates, which is shown in Fig.~\hyperref[fig:fig_6]{\ref*{fig:fig_6}(h)}.
Figure~\ref{fig:fig_7} demonstrates the results of computations for $\textstyle \kappa_a \neq \kappa_b$.
Figure~\hyperref[fig:fig_7]{\ref*{fig:fig_7}(a)} shows that for $\textstyle g_{1a} < g_{2b}$, stronger entanglement is achieved for larger values of $\textstyle \kappa_a$.
The crossover to the regime of $\textstyle g_{1a} > g_{2b}$ results in the opposite behavior of the entanglement entropy when stronger entanglement is achieved for the lower photon leakage rate $\textstyle \kappa_a$, which is shown in Fig.~\hyperref[fig:fig_7]{\ref*{fig:fig_7}(c)}.
Lower photon leakage rate $\textstyle \kappa_b$ gives stronger entanglement, but the dependence of the entanglement entropy on $\textstyle \kappa_b$ is rather weak, especially for $\textstyle g_{1a} > g_{2b}$.
Thus, the results of computations suggest that stronger entanglement of the emitted photons is achieved by increasing the ratios $\textstyle g_{2b}/g_{1a}$ and $\textstyle g_{1b}/g_{2b}$ in the regime of $\textstyle \kappa_a \gtrsim g_{1a}$ and $\textstyle \kappa_b \gtrsim g_{2b}$.

Let us elucidate some features of the obtained behavior of the entanglement entropy using the general considerations discussed in Sec.~\ref{sec:scheme}.
The relaxation of the first transmon not only delivers the photon via resonator $\textstyle A$ to waveguide $\textstyle \alpha$ but also triggers the relaxation of the second transmon leading to the emission of the photon into waveguide $\textstyle \beta$.
For gaining stronger temporal correlations between the emitted photons, we need the photon to be emitted into the waveguide $\textstyle \beta$ shortly after the emission of the photon into waveguide $\textstyle \alpha$ triggered the relaxation of the second transmon.
That is attained by increasing the coupling between the second transmon and resonator $\textstyle B$, i.e., by increasing the ratio $\textstyle g_{2b}/g_{1a}$.
The larger ratios $\textstyle g_{1b}/g_{2b}$ are required for the efficient inhibition of the second transmon relaxation until the moment the first transmon decays, delivering a photon to waveguide $\textstyle \alpha$.
The mechanism of that inhibition is clarified in the second paragraph of Sec.~\ref{sec:scheme}.

\section{Discussion and Summary} \label{sec:concl}
\begin{figure*}[t!] 
	\centering
	\includegraphics{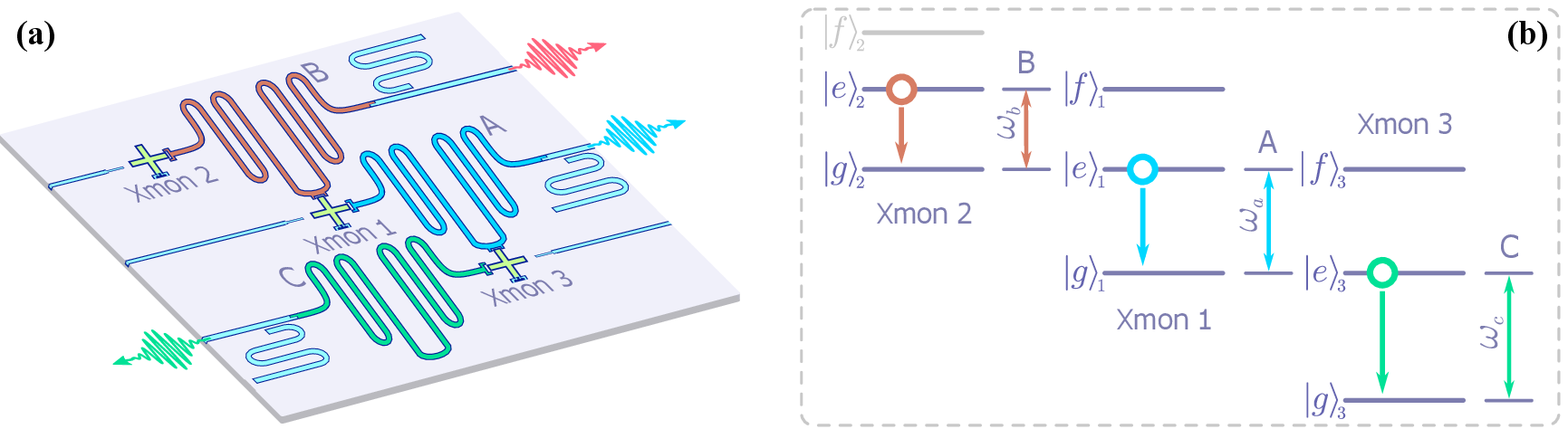}
	\caption{(a) The variant of extension of the proposed scheme of generation of TF entangled photons: schematics of the potential circuit QED setup for generation of three-photon entangled states.	(b) Scheme illustrating the relationships between the frequencies of the resonators and the transition frequencies of the transmons in the setup shown in Fig.~\hyperref[fig:fig_8]{\ref*{fig:fig_8}(a)}. \label{fig:fig_8}}
\end{figure*}

Having outlined the scheme and operational principle of the on-demand source of microwave TF entangled photon pairs and investigated its performance, let us now briefly discuss the general idea of how the proposed scheme can be extended for the generation of multiphoton entangled states. For this purpose, we consider the setup for the generation of the three-photon entangled states illustrated in Fig.~\ref{fig:fig_8}.
Compared to the original setup for the generation of photon pairs shown in Fig.~\ref{fig:fig_2}, here we added one more resonator (marked as $\textstyle C$) with frequency $\textstyle \omega_c$ coupled to the output transmission line and a third transmon coupled to resonators $\textstyle A$ and $\textstyle C$.
Following the logic of Sec.~\ref{sec:scheme}, the frequencies $\textstyle \omega^{ge}_3$ and $\textstyle \omega^{ef}_3$ of $|g\rangle_3\leftrightarrow|e\rangle_3$ and $\textstyle |e\rangle_3\leftrightarrow|f\rangle_3$ transitions of the third transmon are set that $\textstyle \omega^{ge}_3 = \omega_c$ and $\textstyle \omega^{ef}_3 = \omega_a$.
The interrelation between the frequencies of the resonators and the transition frequencies of the transmons is schematically shown in Fig.~\ref{fig:fig_8}(b).
Such a choice of frequencies enables the resonant excitation exchange between $\textstyle |g\rangle_3 \leftrightarrow |e\rangle_3$ transition and resonator $\textstyle C$ and $\textstyle |e\rangle_3\leftrightarrow|f\rangle_3$ transition and resonator $\textstyle A$, while the excitation exchange between $\textstyle |g\rangle_3 \leftrightarrow |e\rangle_3$ transition and between resonator $\textstyle A$ and $\textstyle |e\rangle_3 \leftrightarrow |f\rangle_3$ transition and resonator $\textstyle C$ is inhibited due to detuning.
The relaxation of the third transmon excited state $\textstyle |e\rangle_3$ decouples $\textstyle |e\rangle_3 \leftrightarrow |f\rangle_3$ transition from resonator $\textstyle A$, triggering the relaxation of the first transmon excited state $\textstyle |e\rangle_1$, which, in turn, triggers the relaxation of the second transmon.
Thus, one may anticipate that the photons are emitted into the corresponding waveguides in TF entangled triples.

To summarize, we proposed a feasible scheme for efficient on-demand generation of pairs of microwave TF entangled photons propagating in separate waveguides and analyzed its performance.
The design and working parameters of the considered setup are readily accessible for the state-of-the-art superconducting circuit QED technologies.
Although we outlined a rather concrete superconducting circuit QED implementation of an on-demand source of microwave TF entangled photon pairs, we should stress that the general principle of operation does not rely on the details of the circuit QED realization and can be applied to other physical systems.
We focused our analysis on the pulsed on-demand regime of photon pairs generation.
Therefore, the consideration of the continuous regime of operation is of interest for future work.
Besides, a detailed investigation of the extended schemes (as shown in Fig.~\ref{fig:fig_8}) for the generation of multiphoton TF entangled states constitutes a potential research direction as well.

\begin{acknowledgements}
 The author thanks Andrii Semenov and Andrii Sokolov for useful comments.
 This work was supported by the National Academy of Sciences of Ukraine through the Program of Postdoctoral Researches.
\end{acknowledgements}
\appendix
\section{Pair of 2LEs coupled to resonator} \label{sec:appa}

The paradigmatic system described in the first paragraph of Sec.~\ref{sec:scheme} and schematically shown in Fig.~\hyperref[fig:fig_1]{\ref*{fig:fig_1}(a)} is modeled by the Hamiltonian as follows
\begin{equation} \label{eq:ham_rqq} 
	\begin{split}
	 \opH = & \, \omega c^\dag c + \sum_{j=1}^2 (\omega + \Delta_j) \sigma^+_j \sigma^-_j + \sum_{j=1}^2 g_j (c^\dag\sigma^-_j + \sigma^+_j c) \\
	 & \, + \int^\infty_0 \dd\nu \nu A^\dag_\nu A_\nu + \int \dd\nu f(\nu) (A^\dag_{\nu} c + c^\dag A_\nu).
	\end{split}
\end{equation}
The first three terms in the above Hamiltonian describe the single-mode resonator with frequency $\textstyle \omega$ coupled to a pair of 2LEs, where $\textstyle \Delta_j$ stands for the detuning between the frequencies of the $\textstyle j$-th 2LE and the resonator, parameter $\textstyle g_j$ is the coupling strength of the resonator to the $\textstyle j$-th 2LE ($\textstyle j\in\{1,2\}$).
Operator $\textstyle c$ ($\textstyle c^\dag$) annihilates (creates) a photon in the resonator, and $\textstyle \sigma^{+}_{j}$ ($\textstyle \sigma^{-}_{j}$) rises (lowers) the state of the $\textstyle j$-th 2LE.
The last pair of terms in the Hamiltonian in Eq.~\eqref{eq:ham_rqq} describes the waveguide, represented by a bath of independent bosonic modes, and its coupling to the resonator with strength $\textstyle f(\nu)$.
Operator $\textstyle A_\nu$ ($\textstyle A^\dag_\nu$) annihilates (creates) a photon with frequency $\textstyle \nu$ propagating in the waveguide.

Now, let us demonstrate that in the single-excitation case, the paradigmatic system composed of a pair of 2LEs coupled to the resonator, considered in Sec.~\ref{sec:scheme} and illustrated in Fig.~\hyperref[fig:fig_1]{\ref*{fig:fig_1}(a)}, can be equivalently represented as a 2LE coupled to the $\textstyle \textsf{V}$-configuration 3LE.
To proceed, we formally represent the Hamiltonian $\textstyle \opH$ given by Eq.~\eqref{eq:ham_rqq} in the eigenbasis of the 2LE-resonator Hamiltonian $\textstyle \opH_\mathrm{JC}$. The latter is constituted by the terms describing the resonator, the first 2LE, and their interaction in Eq.~\eqref{eq:ham_rqq}, which together form the Hamiltonian of the Jaynes-Cummings (JC) model \cite{shore1993}.
The JC system ground state $|\gnd\rangle = |0\rangle_\res |g\rangle_1$ is a state with a vacuum field in the resonator $\textstyle |0\rangle_\res$ and the first 2LE in the ground state $\textstyle |g\rangle_1$.
The excited eigenstates of $\opH_\mathrm{JC}$ are the superpositions of the state $\textstyle |n\rangle_\res |g\rangle_1$ containing $\textstyle n$ photons in the resonator with the ground-state 2LE and the state $\textstyle |n-1\rangle_\res |e\rangle_1$ with $\textstyle n-1$ photons in the resonator and the excited-state 2LE \cite{shore1993, blais2004}:
\begin{equation*} \label{eq:jc_eigst}
	\left(
	\begin{array}{c}
		|\pol_n^-\rangle \\
		|\pol_n^+\rangle
	\end{array}
	\right)
	=
	\left(
	\begin{array}{cc}
		\cos \mu_n & - \sin \mu_n \\
		\sin \mu_n & \cos \mu_n
	\end{array}
	\right)
	\left(
	\begin{array}{c}
		|n\rangle_\res |g\rangle_1 \\
		|n-1\rangle_\res |e\rangle_1
	\end{array}
	\right),
\end{equation*}
where
\begin{equation*}
	\tan \mu_n = \sqrt{\frac{\Lambda_n - \Delta_1}{\Lambda_n+\Delta_1}}, \quad \Lambda_n = \sqrt{4 n g^2 + \Delta_1^2}.
\end{equation*}
The JC eigenstates $\textstyle |\pol^\pm_n\rangle$ correspond to the eigenfrequencies $\textstyle E^\pm_n = n \omega
+ (\Delta_1 \pm \Lambda_n)/2$.

Restricting ourselves to the resonant regime of the 2LE-resonator coupling ($\textstyle \Delta_1=\Delta_2=0$) and the single-excitation domain, as considered in Sec.~\ref{sec:scheme}, we project the system Hamiltonian $\textstyle \opH$ on the lowest eigenstates $\textstyle |\gnd\rangle$ and $\textstyle |\pol^\pm_1\rangle$ of the JC Hamiltonian.
Thus, one has $\textstyle \opH \rightarrow \varPi \opH \varPi$, where $\textstyle \varPi = |\gnd\rangle\langle\gnd| + \sum_{\pm}|\pol^\pm_1\rangle\langle\pol^\pm_1|$, which gives:
 \begin{equation*} \label{eq:ham_2}
 	\begin{split}
 		\opH = & \, \omega \sigma^+_2\sigma^-_2 + \sum_{\pm} (\omega \pm g_1) |\pol^\pm\rangle \langle\pol^\pm|
 		+ \int^\infty_0 \dd\nu \nu A^\dag_\nu A_\nu \\
 		& \, + \frac{g_2}{\sqrt{2}}\sum_{\pm} \left(\sigma^+_2 |\gnd\rangle\langle\pol^\pm| + |\pol^\pm\rangle\langle\gnd| \sigma^-_2\right) \\
 		& \, + \int^\infty_0\dd\nu \, \frac{f(\nu)}{\sqrt{2}} \sum_{\pm}(A^\dag_{\nu} |\gnd\rangle\langle\pol^\pm| + |\pol^\pm\rangle\langle\gnd| A_\nu),
 	\end{split}
 \end{equation*}
where for brevity we dropped the subscripts indicating the photon number in the notations of JC eigenstates.
The above Hamiltonian describes the 3LE coupled to the 2LE and the waveguide.
The 3LE is constituted by the ground state $|\gnd\rangle$ and a pair of excited states $|\pol^\pm\rangle$.
The frequency of $\textstyle |\gnd\rangle \leftrightarrow |\pol^\pm\rangle$ transition is $\omega\pm g_1$.
The levels of the 3LE are arranged in the $\mathsf{V}$-type configuration.
Each transition of this $\mathsf{V}$3LE is coupled to the second 2LE with strength $g_2/\sqrt{2}$ and to the waveguide with strength $f(\nu)/\sqrt{2}$.
This equivalent representation of the system composed of the resonator resonantly coupled to a pair of 2LEs is illustrated in Fig.~\hyperref[fig:fig_1]{\ref*{fig:fig_1}(b)}.

\subsection*{Dynamics of the resonator-2LEs system}

\begin{figure}[t!]
	\centering
	\includegraphics{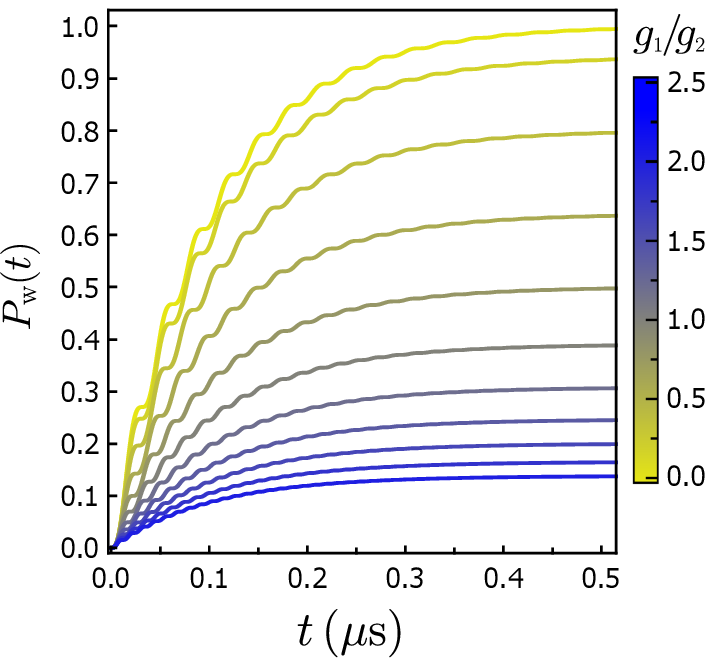}
	\caption{Effect of the 2LEs couplings ratio $\textstyle g_1/g_2$ on the evolution of the waveguide population $\textstyle P_\mathrm{w}(t)$ in the system shown in Fig.~\hyperref[fig:fig_1]{\ref*{fig:fig_1}(a)}.
		The parameters used for calculations are as follows: $\textstyle g_2/2\pi = 10\,\mathrm{MHz}$ and $\textstyle \kappa/2\pi = 2\,\mathrm{MHz}$. \label{fig:fig_9}}
\end{figure}

The waveguide population at time $\textstyle t$ is given by:
\begin{equation} \label{eq:wg_pop}
 P_\mathrm{w}(t) = \kappa\int^t_0 \dd\tau \langle c^\dag(\tau) c(\tau) \rangle,
\end{equation}
with $\textstyle \kappa = 2\pi f^2(\omega)$ being the resonator decay rate into the waveguide.
The averaging in Eq.~\eqref{eq:wg_pop} goes over the initial state of the system $\textstyle |\psi_0\rangle = \sigma^+_2|0\rangle$ with $\textstyle |0\rangle=|0\rangle_\res|g\rangle_1|g\rangle_2$ being the vacuum state of the resonator-2LEs system.
In the single-excitation case, the resonator population $\textstyle \langle c^\dag(t) c(t) \rangle$ can be evaluated using the relation
$\textstyle \langle c^\dag(t) c(t) \rangle = \langle\psi_0|c^\dag(t)|0\rangle\langle 0|c(t)|\psi_0\rangle$.
The evolution of the matrix element $\textstyle \langle 0|c(t)|\psi_0\rangle$ is governed by the set of equations of motion:
 \begin{equation*}
    \pdt \!\left(\!
    \begin{array}{c}
	\langle 0|c(t)|\psi_0\rangle \\
	\langle 0|\sigma^-_1(t)|\psi_0\rangle \\
	\langle 0|\sigma^-_2(t)|\psi_0\rangle
   \end{array}
   \!\right)
 	= -\ii 
 	\begin{pmatrix}
 		\widetilde{\omega} & g_1 & g_2 \\
 		g_1 & \omega & 0 \\
 		g_2 & 0 & \omega
 	\end{pmatrix}
    \!\!
    \left(\!
    \begin{array}{c}
     \langle 0|c(t)|\psi_0\rangle \\
     \langle 0|\sigma^-_1(t)|\psi_0\rangle \\
     \langle 0|\sigma^-_2(t)|\psi_0\rangle
    \end{array}
    \!\right),
 \end{equation*}
where $\textstyle \widetilde{\omega} = \omega - \ii\kappa/2$.

Using the solution of the above set of equations in Eq.~\eqref{eq:wg_pop}, one obtains
\begin{equation} \label{eq:Pwt}
 P_\mathrm{w}(t) = \frac{g_2^2}{g_1^2+g_2^2} \Bigg[1-F(t)\exp\left(-\frac{\kappa}{2}t\right)\Bigg],
\end{equation}
with $\textstyle F(t)$ given by
\begin{equation*}
	F(t) = 1+\kappa \sin (\varOmega t) \, \frac{\kappa \sin (\varOmega t) + 4\varOmega \cos (\varOmega t)}{8\varOmega^2},
\end{equation*}
where $\textstyle \varOmega = \sqrt{g_1^2 + g_2^2 - (\kappa/4)^2}$.

Figure~\ref{fig:fig_9} shows the dependence of the dynamics of the waveguide population $\textstyle P_\wg(t)$ on the ratio of the couplings $\textstyle g_1/g_2$.
Calculations demonstrate that the increase of $\textstyle g_1/g_2$ leads to the inhibition of the photon leakage into the waveguide.
It follows from Eq.~\eqref{eq:Pwt} that $\textstyle P_\mathrm{w}(t \rightarrow \infty)<1$ for $\textstyle g_1\neq 0$.
For $\textstyle g_1 \gg g_2$, one has $\textstyle P_\mathrm{w}(t \rightarrow \infty)\ll 1$ implying that the leakage into the waveguide is suppressed and the photon remains trapped within the resonator-2LEs system.

\begin{widetext}
\section{Derivation of the effective Hamiltonian} \label{sec:appb}
\subsection{Operators in the dressed basis} \label{sec:dress_ops}
Using the definition of the unitary operator $\textstyle \opU$ given by Eq.~\eqref{eq:Udisp} along with the Baker-Campbell-Hausdorff formula, one obtains the following expressions for the resonator photon annihilation operators in the dressed basis:
\begin{subequations} \label{eq:ops_ab}
	\begin{equation}
	 \opU^\dag a \, \opU \approx \left(1 + \frac{\lambda_{1a}^2}{2} \opZ^{fe}_1\right) a + \lambda_{1a} \sigma_1^{ef}
	 - \frac{\lambda_{1a} \lambda_{1b}}{2} b^\dag \sigma_1^{gf},
	\end{equation}
    \begin{equation}
     \opU^\dag b \, \opU \approx \left(1 + \frac{\lambda_{1b}^2}{2} \opZ^{eg}_1 + \frac{\lambda_{2b}^2}{2} \opZ^{fe}_2\right) b
     + \lambda_{1b} \sigma_1^{ge} + \lambda_{2b} \sigma_2^{ef} + \frac{\lambda_{1a} \lambda_{1b}}{2} a^\dag \sigma_1^{gf},
    \end{equation}
\end{subequations}
where we keep the terms up to the second order in the small parameters $\lambda_{1a}$, $\lambda_{1b}$, and $\lambda_{2b}$.
The resonator photon number operators $\textstyle \NA$ and $\textstyle \NB$ in the new basis read as
\begin{subequations} \label{eq:ops_aabb}
	\begin{equation}
		\opU^\dag \NA \, \opU \approx \left(1 + \lambda_{1a}^2 \opZ^{fe}_1\right) \NA + \lambda_{1a} \left(a^\dag \sigma_1^{ef} + \sigma_1^{fe} a\right) + \lambda^2_{1a} \sigma_1^{ff}
		- \frac{\lambda_{1a} \lambda_{1b}}{2} \left(a^\dag b^\dag \sigma_1^{gf} + \sigma_1^{fg} a b\right),
	\end{equation}
	\begin{equation}
		\begin{split}
		  \opU^\dag \NB \, \opU \approx & \, \left(1 + \lambda_{1b}^2 \opZ^{eg}_1 + \lambda_{2b}^2 \opZ^{fe}_2\right) \NB
		  + \lambda_{1b} \left(b^\dag \sigma_1^{ge} + \sigma_1^{eg} b\right)
		  + \lambda_{2b} \left(b^\dag \sigma_2^{ef} + \sigma_2^{fe} b\right) \\
		  & \, + \lambda_{1b}^2 \sigma_1^{ee} + \lambda_{2b}^2 \sigma_2^{ff}
		  + \frac{\lambda_{1a} \lambda_{1b}}{2} \left(a^\dag b^\dag \sigma_1^{gf} + \sigma_1^{fg} a b\right)
		  + \lambda_{1b} \lambda_{2b} \left(\sigma_2^{fe} \sigma_1^{ge} + \sigma_1^{eg} \sigma_2^{ef}\right).
		\end{split}
	\end{equation}
\end{subequations}
In the dressed basis, the transmon ladder operators $\textstyle \sigma^{ge}_{1,2}$ and $\textstyle \sigma^{ef}_{1,2}$ acquire the form as follows
\begin{subequations} \label{eq:ops_lad}
\begin{equation}
	\begin{split}
		\opU^\dag \sigma_1^{ge} \opU \approx & \left[1 - \frac{\lambda_{1a}^2}{2} \NA - \frac{\lambda_{1b}^2}{2}(2\NB + 1)\right]\sigma_1^{ge}
		+ \lambda_{1b} \opZ^{eg}_1 b + \lambda_{1a} a^\dag \sigma_1^{gf} \\
		& \, - \lambda^2_{1b} \sigma_1^{eg} b^2 + \lambda_{1a} \lambda_{1b} a^\dag b \sigma_1^{ef}
		+ \frac{\lambda_{1a} \lambda_{1b}}{2} \sigma_1^{fe} a b  + \frac{\lambda_{1b} \lambda_{2b}}{2} \opZ^{eg}_1 \sigma_2^{ef},
	\end{split}
\end{equation}
\begin{equation}
	\begin{split}
		\opU^\dag \sigma_1^{ef} \opU \approx & \left[1 - \frac{\lambda_{1a}^2}{2}(2\NA + 1) -\frac{\lambda^2_{1b}}{2}(\NB + 1)\right]\sigma_1^{ef} + \lambda_{1a} \opZ^{fe}_1 a - \lambda_{1b} b^\dag \sigma_1^{gf} \\
		& \, + \lambda_{1a}\lambda_{1b} \, b^\dag a \, \sigma_1^{ge} + \frac{\lambda_{1a} \lambda_{1b}}{2} \, a b \, \sigma_1^{eg} - \lambda_{1a}^2 \sigma_1^{fe} a^2 - \frac{\lambda_{1b} \lambda_{2b}}{2}\sigma_1^{gf}\sigma_2^{fe},
	\end{split}
\end{equation}
\begin{equation}
	\opU^\dag \sigma_2^{ge} \opU \approx \left(1 - \frac{\lambda_{2b}^2}{2} \NB\right)\sigma_2^{ge} + \lambda_{2b} b^\dag \sigma_2^{gf} + \frac{\lambda_{1b} \lambda_{2b}}{2} \sigma_1^{eg} \sigma_2^{gf}
\end{equation}
\begin{equation}
	\opU^\dag \sigma_2^{ef} \opU \approx \left[1 - \frac{\lambda_{2b}^2}{2} (2\NB + 1)\right] \sigma_2^{ef} + \lambda_{2b} \opZ^{fe}_2 b - \lambda_{2b}^2 \sigma_2^{fe} b^2 + \frac{\lambda_{1b} \lambda_{2b}}{2} \opZ^{fe}_2 \sigma_1^{ge}
\end{equation}
\end{subequations}
For the projection operators on the transmons eigenstates $\textstyle \sigma_{1,2}^{ee}$ and $\textstyle \sigma_{1,2}^{ff}$, one has:
\begin{subequations} \label{eq:ops_prj}
\begin{equation}
	\begin{split}
		\opU^\dag \sigma_1^{ee} \opU \approx & \left(1 - \lambda_{1b}^2\right) \sigma_1^{ee}
		+ \lambda_{1a} \left(a^\dag \sigma_1^{ef} + \sigma_1^{fe} a\right)
		- \lambda_{1b} \left(b^\dag \sigma_1^{ge} + \sigma_1^{eg} b\right)
		+ \lambda_{1a}^2 \sigma_1^{ff} \\
		& \, + \lambda_{1a}^2 \opZ^{fe}_1 \NA - \lambda_{1b}^2 \opZ^{eg}_1 \NB
		- \lambda_{1a} \lambda_{1b} \left(a^\dag b^\dag \sigma_1^{gf} + \sigma_1^{fg} a b\right)
		- \frac{\lambda_{1b} \lambda_{2b}}{2}\left(\sigma_2^{fe}\sigma_1^{ge} + \sigma_1^{eg}\sigma_2^{ef}\right),
	\end{split}
\end{equation}
\begin{equation}
	\begin{split}
		\opU^\dag \sigma_1^{ff} \opU \approx \left(1 - \lambda_{1a}^2\right) \sigma_1^{ff}
		- \lambda_{1a} \left(a^\dag \sigma_1^{ef} + \sigma_1^{fe} a\right)	- \lambda_{1a}^2 \opZ^{fe}_1 \NA
		+ \frac{\lambda_{1a} \lambda_{1b}}{2}\left(a^\dag b^\dag \sigma_1^{gf} + \sigma_1^{fg} a b\right),
	\end{split}
\end{equation}
\begin{equation}
	\opU^\dag \sigma_2^{ee} \opU \approx \sigma_2^{ee} + \lambda_{2b} \left[b^\dag \sigma_2^{ef} + \sigma_2^{fe} b\right]
	+ \lambda_{2b}^2 \opZ^{fe}_2 \NB + \lambda_{2b}^2 \sigma_2^{ff}
	+ \frac{\lambda_{1b} \lambda_{2b}}{2}\left(\sigma_2^{fe}\sigma_1^{ge} + \sigma_1^{eg}\sigma_2^{ef}\right),
\end{equation}
\begin{equation}
	\opU^\dag \sigma_2^{ff} \opU \approx (1-\lambda_{2b}^2) \sigma_2^{ff}
	- \lambda_{2b} \left(\sigma_2^{fe} b + b^\dag \sigma_2^{ef}\right)
	- \lambda_{2b}^2 \opZ^{fe}_2 \NB
	- \frac{\lambda_{1b} \lambda_{2b}}{2} \left(\sigma_2^{fe} \sigma_1^{ge} + \sigma_1^{eg} \sigma_2^{ef}\right).
\end{equation}
\end{subequations}


Using Eqs.~\eqref{eq:ops_ab}--\eqref{eq:ops_prj} and keeping the terms up to the first order in small parameters $\textstyle \lambda_{1a}$, $\textstyle \lambda_{1b}$, and $\textstyle \lambda_{2b}$, one arrives at the Hamiltonian given by Eq.~\eqref{eq:ham_eff}.

\subsection{Role of Purcell filters} \label{sec:purcell}
Note that in the dressed basis the Hamiltonian of the resonator-waveguide couplings $\textstyle \opH_\mathrm{r-w}$ acquires the form: $\textstyle \opU^\dag \opH_{\res-\wg} \opU \approx \opH_{\res-\wg} + \opH_{\tmn-\wg}$, where $\textstyle \opH_{\res-\wg}$ is given by Eq.~\eqref{eq:ham_rw}, and the term $\textstyle \opH_{\tmn-\wg}$ reads as
\begin{equation} \label{eq:ham_qw}
	\opH_{\tmn-\wg} = \lambda_{1a} \int^\infty_0 \dd\nu f_{a}(\nu) \big(\alpha^\dag_\nu \sigma_1^{ef} + \sigma_1^{fe} \alpha_\nu\big)
	 + \lambda_{1b} \int^\infty_0\dd\nu f_{b}(\nu)  \big(\beta^\dag_\nu \sigma_1^{ge} + \sigma_1^{eg}\beta_\nu\big)
	 + \lambda_{2b} \int^\infty_0\dd\nu f_{b}(\nu) \big(\beta^\dag_\nu \sigma_2^{ef} + \sigma_2^{fe}\beta_\nu\big).
\end{equation}
The first pair of terms in $\textstyle \opH_{\tmn-\wg}$ describes the coupling between $\textstyle |e\rangle_1 \leftrightarrow |f\rangle_1$ and $\textstyle |g\rangle_1 \leftrightarrow |e\rangle_1$ transitions of the first transmon and waveguides $\textstyle \alpha$ and $\textstyle \beta$, respectively.
The third term in Eq.~\eqref{eq:ham_qw} describes the direct coupling between $\textstyle |e\rangle_2 \leftrightarrow |f\rangle_2$ transition of the second transmon and waveguide $\textstyle \beta$ leading to the relaxation of the state $\textstyle |f\rangle_2$.
These direct couplings of the transmon transitions to the waveguides lead to that both excitations can be emitted into one waveguide instead of being emitted into the separate waveguides.
For mitigating this superfluous emission processes, we use the Purcell filters.
In the proposed setup (see Fig.~\ref{fig:fig_2}), the latter are represented by the resonators side-coupled to the output waveguides.
The frequencies of these resonators correspond to the dressed frequencies $\textstyle \bar{\omega}_1^{ge}$ and $\textstyle \bar{\omega}_1^{ef}$ of $\textstyle |g\rangle_1 \leftrightarrow |e\rangle_1$ and $\textstyle |e\rangle_1 \leftrightarrow |f\rangle_1$ transitions.
The filter resonators reject the radiation in the narrow bands of frequencies in the vicinity of $\textstyle \bar{\omega}_1^{ge}$ and $\textstyle \bar{\omega}_1^{ef}$ and transmit the radiation with frequencies outside these bands \cite{reed2010}.
Thus, the Purcell filters inhibit the unwanted emission channels of the first transmon described by the first two terms in the Hamiltonian~\eqref{eq:ham_qw}.
As we demonstrate in Sec.~\ref{sec:evol}, no more than one photon could reside in either of the resonators, so $\textstyle |e\rangle_2 \rightarrow |f\rangle_2$ transition of the second transmon is not involved in the dynamics of the system, and $\textstyle |f\rangle_2$ level is not excited.
Thus, the Purcell filter cutting the emission from this transition into waveguide $\textstyle \beta$ is not required.
Since the processes of direct relaxation of transmons into the waveguides are essentially suppressed by the Purcell filters and occur on the timescales much longer than the photon emission times, for simplicity, in the Hamiltonian $\textstyle \Heff$, we drop the term $\textstyle \opH_{\tmn-\wg}$ describing these processes. 

\section{Derivation of evolution equations} \label{sec:appd}
For deriving Eq.~\eqref{eq:eqmot} governing the evolution of the probability amplitudes, we follow the lines of Appendix B in Ref.~\cite{sto2019}.
For this purpose, we start with the derivation of the equations of motion for the operators.

The effective Hamiltonian~\eqref{eq:ham_eff} generates the equations of motion for the waveguide variables $\textstyle \alpha_\nu(t)$ and $\textstyle \beta_\nu(t)$ as follows:
\begin{subequations}
	\begin{equation}
		\pdt \alpha_\nu(t) = - \ii \nu \alpha_\nu(t) - \ii f_a(\nu) a(t),
	\end{equation}
	\begin{equation}
		\pdt \beta_\nu(t) = - \ii \nu \beta_\nu(t) - \ii f_b(\nu) b(t),
	\end{equation}
\end{subequations}
with the formal solutions written as
\begin{subequations}
	\begin{equation} \label{eq:sln_alpha}
		\alpha_\nu(t) = \tilde{\alpha}_\nu(t) - \ii f_a(\nu) \int^t_0 \dd \tau \ee^{-\ii \nu (t-\tau)} a(\tau),
	\end{equation}
	\begin{equation} \label{eq:sln_beta}
		\beta_\nu(t) = \tilde{\beta}_\nu(t) - \ii f_b(\nu) \int^t_0 \dd \tau \ee^{-\ii \nu (t-\tau)} b(\tau),
	\end{equation}
\end{subequations}
where $\textstyle \tilde{\alpha}_\nu(t) = \alpha_\nu(0)\ee^{-\ii \nu t}$ and $\textstyle \tilde{\beta}_\nu(t) = \beta_\nu(0)\ee^{-\ii \nu t}$ stand for the annihilation operators of a photon propagating as a free excitation in waveguide $\textstyle \alpha$ and $\textstyle \beta$, respectively.

The equation of motion for the annihilation operator $\textstyle a(t)$ of the photon in resonator $\textstyle A$ reads as
%
\begin{equation} \label{eq:eq1_a}
	\pdt a(t)   = - \ii \left[\omegaa + \chi_{1a} \opZ_1^{fe}(t)\right] a(t) - \ii g_{1a} \sigma_1^{ge}(t)
	- \ii \varUpsilon b^\dag(t) \sigma_1^{gf}(t) - \ii \int^\infty_0 \dd\nu f_a(\nu) \alpha_\nu(t).
\end{equation}
Plugging Eq.~\eqref{eq:sln_alpha} into the last term on the right-hand side of the above equation gives
\begin{equation} \label{eq:int_fa}
	\begin{split}
		\int^\infty_0 \dd\nu f_a(\nu) \alpha_\nu(t) & = \alpha_\mathrm{in}(t) - \ii \int^t_0 \dd\tau \int^\infty_0 \dd\nu f^2_a(\nu) \ee^{-\ii \nu (t-\tau)} a(\tau),
	\end{split}
\end{equation}
where $\textstyle \alpha_\mathrm{in}(t) = \int^\infty_0 \dd\nu \, f_a(\nu) \widetilde{\alpha}_{\nu}(t)$.
Using Eq.~\eqref{eq:eq1_a} and that $\textstyle |\chi_{1a}|\ll\omegaa$, one can represent the operator $\textstyle a(t)$ as $\textstyle a(t) = \mathfrak{a}(t)\ee^{-\ii\omegaa t}$, where $\textstyle \mathfrak{a}(t)$ stands for the slowly-varying part of $a(t)$.
With such a representation, one can notice that only the frequencies in the vicinity of $\textstyle \omegaa$ contribute significantly to the integral over $\textstyle \tau$ on the right-hand side of Eq.~\eqref{eq:int_fa}.
Thus, one can extend the lower bound of integration over $\textstyle \nu$ to $\textstyle -\infty$, and neglect the frequency dependence of the coupling parameter $\textstyle f_a(\nu)$ assuming that $\textstyle f_a(\nu) \approx f_a(\omegaa)$.
Using these approximations, one obtains
\begin{equation} \label{eq:int_fa2}
	\int^\infty_0 \dd\nu f_a(\nu) \alpha_\nu(t) \approx \alpha_\mathrm{in}(t) - \ii \frac{\kappa_a}{2} a(t), \quad \kappa_a = 2\pi f^2_{a}(\omega_a).
\end{equation}
Substituting this result into Eq.~\eqref{eq:eq1_a} yields
\begin{equation} \label{eq:eq2_a}
	\pdt a(t) = - \ii \left[\omegaa - \ii \frac{\kappa_a}{2} + \chi_{1a} \opZ_1^{fe}(t)\right] a(t)
	- \ii g_{1a} \sigma_1^{ge}(t) - \ii \varUpsilon b^\dag \sigma_1^{gf}(t) - \ii \alpha_\mathrm{in}(t).
\end{equation}
The equation of motion for the operator $b(t)$ annihilating a photon in the resonator $B$ is obtained similarly to Eq.~\eqref{eq:eq2_a} and reads as
\begin{equation} \label{eq:eq_b}
	\pdt b(t) = - \ii \left[\omegab - \ii\frac{\kappa_b}{2} + \chi_{1b} \opZ_1^{eg}(t) + \chi_{2b} \opZ_2^{fe}(t)\right] b(t)
	-\ii g_{1b} \sigma_1^{ef}(t) - \ii g_{2b} \sigma_2^{ge}(t)
	- \ii \varUpsilon a^\dag(t) \sigma_1^{gf}(t)
	- \ii \beta_\mathrm{in}(t),
\end{equation}
where $\textstyle \beta_\mathrm{in}(t) = \int^\infty_0\dd\nu \, f_b(\nu) \tilde{\beta}_{\nu}(t)$.
For the derivation of Eq.~\eqref{eq:eq_b}, we used the approximate relation
\begin{equation} \label{eq:int_fb}
	\int^\infty_0\dd\nu f_b(\nu)\beta_\nu(t) \approx \beta_\mathrm{in}(t) - \ii \frac{\kappa_b}{2} b(t), \quad \kappa_b = 2\pi f^2_b(\omegab),
\end{equation}
which is obtained by analogy to Eq.~\eqref{eq:int_fa2}.

The equations of motion for the required ladder operators of the transmons read as follows:
\begin{subequations}
	\begin{equation}
		\pdt \sigma_1^{ge}(t) = - \ii\left[\omega_1^{ge} + \chi_{1b} - \chi_{1a} \NA(t) + 2\chi_{1b} \NB(t)\right] \sigma_1^{ge}(t) + \ii g_{1a} \opZ_1^{eg}(t) a(t) -\ii g_{1b} b^\dag(t) \sigma_1^{gf}(t)
		+ \ii \varUpsilon \sigma_1^{fe}(t) a(t) b(t),
	\end{equation}
	\begin{equation}
		\pdt \sigma_1^{gf}(t) =  - \ii \left[\omega_1^{ge} + \omega_1^{ef} + \chi_{1a}\right]\sigma_1^{gf}(t) - \ii g_{1a} \sigma_1^{ef}(t) a - \ii g_{1b} \sigma_1^{ge}(t) b(t) + \ii \varUpsilon \opZ_1^{fg}(t) a(t) b(t),
	\end{equation}
	\begin{equation}
		\pdt \sigma_2^{ge}(t) = - \ii \left[\omega_2^{ge} - \chi_{2b} \NB(t)\right] \sigma_2^{ge}(t) + \ii g_{2b} \opZ_2^{eg}(t) b(t).
	\end{equation}
\end{subequations}


Using Eq.~\eqref{eq:wavefunc}, one can express $\textstyle \varPhi_{\nu,\nu'}(t)$ as $\textstyle \varPhi_{\nu,\nu'}(t) = \vacbra \alpha_\nu \beta_{\nu'}|\varPsi(t)\rangle$.
Then, we use the standard quantum-mechanical relations $\textstyle |\varPsi(t)\rangle = \ee^{-\ii \Heff t}|\varPsi(0)\rangle$ and $\textstyle \mathcal{O}(t) = \ee^{\ii \mathcal{H}_\mathrm{eff} t} \mathcal{O} \ee^{-\ii \mathcal{H}_\mathrm{eff} t}$, where $\textstyle \opO$ and $\textstyle \opO(t)$ denote the operators in the Schr\"{o}dinger and Heisenberg representations, correspondingly.
This leads us to the expression $\textstyle \varPhi_{\nu,\nu'}(t) = \vacbra \alpha_\nu(t) \beta_{\nu'}(t) |\Psi_\mathrm{in}\rangle$, where we used that $\textstyle |\varPsi(0)\rangle \equiv |\Psi_\mathrm{in}\rangle$ and $\textstyle \ee^{-\ii \mathcal{H}_\mathrm{eff} t}|\varnothing\rangle = |\varnothing\rangle$.
Using such an approach for the rest of the probability amplitudes entering the state vector given by Eq.~\eqref{eq:wavefunc}, one obtains:
\begin{equation} \label{eq:melem}
	\boldsymbol{\mu}(t) =
	\left[
	\begin{array}{c}
		\varPhi_{\nu,\nu'}(t) \\
		\varXi^{\alpha}_\nu(t) \\
		\varXi^{\beta}_\nu(t) \\
		\varTheta^{\alpha}_\nu(t) \\
		\varTheta^{\beta}_\nu(t) \\
		R(t) \\
		S(t) \\
		W^a(t) \\
		W^b(t) \\
		X(t)
	\end{array}
	\right]
	=
	\big\langle \varnothing\big|\!
	\left[\!
	\begin{array}{c}
		\alpha_{\nu}(t) \beta _{\nu'}(t) \\
		{\alpha}_{\nu}(t) {b}(t) \\
		{\beta}_{\nu}(t) {a}(t) \\
		{\alpha}_{\nu}(t) {\sigma}_2^{ge}(t) \\
		{\beta}_{\nu}(t) {\sigma}_2^{ge}(t) \\
		{a}(t) {b}(t) \\
		{\sigma}_1^{gf}(t) \\
		{a}(t) {\sigma}_2^{ge}(t) \\
		{b}(t) {\sigma}_1^{ge}(t) \\
		{\sigma}_1^{ge}(t) {\sigma}_2^{ge}(t)
	\end{array}
	\!\right]\!
	\big|\Psi_\mathrm{in}\big\rangle.
\end{equation}
Then, using the equations of motion for the corresponding operators derived above, we derive the set of evolution equations for the matrix elements standing on the right-hand side of Eq.~\eqref{eq:melem}.
This leads us to the set of evolution equations for the probability amplitudes given by Eq.~\eqref{eq:eqmot}, where we used the narrowband approximation assuming that $\textstyle f_a(\nu)\approx f_a(\omegaa)$ and $\textstyle f_b(\nu)\approx f_b(\omegab)$.
\end{widetext}

\bibliography{bibliography}
\end{document}